\documentclass[11pt]{article}
\title{Selection of Ultrahigh-Dimensional Interactions Using $L_0$ Penalty}
\author{Tonglin Zhang\footnote{Department of Statistics, Purdue University, 150 North University Street, West Lafayette, IN 47907-2067, Email: tlzhang@purdue.edu}}
\usepackage{subfigure}
\usepackage{epsfig}
\usepackage{mathrsfs}
\usepackage{amsfonts}
\usepackage{amsmath}
\usepackage{wrapfig,lipsum,booktabs}
\usepackage{bm}
\usepackage{algorithm}
\usepackage{algpseudocode}
\usepackage{pifont}
\def\qed{\hfill$\diamondsuit$}

\newtheorem{thm}{Theorem}
\newtheorem{cor}{Corollary}
\newtheorem{lem}{Lemma}

\begin{document}
\maketitle
\def\eqalign#1{\null\,\vcenter{\openup\jot\ialign
              {\strut\hfil$\displaystyle{##}$&$\displaystyle{{}##}$
               \hfil\crcr#1\crcr}}\,}


\setcounter{page}{1}
\begin{abstract}

Selecting interactions from an ultrahigh-dimensional statistical model with $n$ observations and $p$ variables when $p\gg n$ is difficult because the number of candidates for interactions is $p(p-1)/2$ and a selected model should satisfy the strong hierarchical (SH) restriction. A new method called the SHL0 is proposed to overcome the difficulty. The objective function of the SHL0 method is composed of a loglikelihood function and an $L_0$ penalty. A well-known approach in theoretical computer science called local combinatorial optimization is used to optimize the objective function. We show that any local solution of the SHL0 is consistent and enjoys the oracle properties, implying that it is unnecessary to use a global solution in practice. Three additional advantages are: a tuning parameter is used to penalize the main effects and interactions; a closed-form expression can derive the tuning parameter; and the idea can be extended to arbitrary ultrahigh-dimensional statistical models. The proposed method is more flexible than the previous methods for selecting interactions. A simulation study of the research shows that the proposed SHL0 outperforms its competitors.

\end{abstract}



\section{Introduction}
\label{sec:introduction}

Selecting interactions from an ultrahigh-dimensional statistical model with $n$ observations and $p$ variables when $p\gg n$ is more difficult than the main effects because the number of candidates for the main effects and interactions quadratically increases with $p$ and a selected model should satisfy the strong hierarchical (SH) restriction. For instance, suppose that an ultrahigh-dimensional data set is composed of a response vector ${\bm y}=(y_1,\dots,y_n)^\top\in\mathbb{R}^n$ and an explanatory variable matrix ${\bf X}_{M}=[{\bm x}_1,\dots,{\bm x}_p]\in\mathbb{R}^{n\times{p}}$, where ${\bm x}_j=(x_{1j},\dots,x_{nj})^\top\in\mathbb{R}^n$ satisfying ${\bm 1}^\top{\bm x}_j=0$ and $\|{\bm x}_j\|^2=n$ represents the $j$th explanatory variable. The probability density function (PDF) or the probability mass function (PMF) of ${\bm y}$ has the standard form of the exponential family distribution as
\begin{equation}
\label{eq:distribution of exponential family}
f({\bm y})=\exp\left\{{{\bm y}\circ{\bm\theta}\over\phi}-b({\bm\theta})+c({\bm y},\phi)\right\},
\end{equation}
where $b({\bm\theta})=(b(\theta_1),\dots,b(\theta_n))^\top\in\mathbb{R}^n$ is derived by transformation $b(\cdot)$ on ${\bm\theta}=(\theta_1,\dots,\theta_n)^\top$ for each $i\in\{1,\dots,n\}$, ${\bm y}\circ{\bm\theta}$ is the Hadamard product (i.e., element-wise product) of vectors, $c({\bm y},\phi)\in\mathbb{R}^n$ is a vector of normalized constants, and $\phi\in\mathbb{R}^+$ is a scale dispersion parameter. The exponential family distribution can be specified for normal, binomial, or Poisson distributions. It satisfies $\mu_i={\rm E}(y_i)=b'(\theta_i)$ and ${\rm var}(y_i)= \phi b''(\theta_i)$. In many cases, $\phi$ is not present, implying that there is $\phi=1$ in~\eqref{eq:distribution of exponential family}. Examples include the Poisson or binomial distributions. With a link $g(\cdot)$, a full ultrahigh-dimensional generalized linear model (HDGLM) has the form of
\begin{equation}
\label{eq:high-dimensional GLM with interaction effects}
g({\bm\mu})=g[b'({\bm\theta})]={\bm\eta}={\bf X}{\bm\beta}=\beta_0+\sum_{j=1}^p {\bm x}_j\beta_j+\sum_{j=1}^{p-1}\sum_{k=p+1}^p {\bm x}_j\circ{\bm x}_k\beta_{jk},
\end{equation}
where ${\bf X}=[{\bm x}_0,{\bf X}_M,{\bf X}_I]\in\mathbb{R}^{n\times[1+p(p+1)/2]}$ with ${\bm x}_0={\bm 1}\in\mathbb{R}^n$ and ${\bf X}_I=[{\bm x}_{j}\circ{\bm x}_k: 1\le j<k\le p]\in\mathbb{R}^{n\times[p(p-1)/2]}$ is the design matrix, ${\bm\beta}=(\beta_0,{\bm\beta}_M^\top,{\bm\beta}_I^\top)^\top$ with $\beta_0\in\mathbb{R}$ for the intercept, ${\bm\beta}_M=(\beta_1,\dots,\beta_p)^\top\in\mathbb{R}^{p}$ for the main effects, and ${\bm\beta}_{I}=(\beta_{12},\dots,\beta_{1p},\beta_{23},\dots,\beta_{2p},\dots,\beta_{(p-1)p})^\top\in\mathbb{R}^{p(p-1)/2}$ for the interactions is the parameter vector, ${\bm\mu}=(\mu_1,\dots,\mu_n)^\top=b'({\bm\theta})={\rm E}({\bm y})$ is the mean vector of the response, and ${\bm\eta}=(\eta_1,\dots,\eta_n)^\top$ is the vector of the linear components. 

It is assumed that only a few components of ${\bm\beta}$ are nonzero, implying that we need to select the important main effects and interactions from~\eqref{eq:high-dimensional GLM with interaction effects}. In particular, denote the true parameter vector as ${\bm\beta}^*=(\beta_0^*,{\bm\beta}_M^{*\top},{\bm\beta}_I^{*\top})^\top$ with $\beta_0^*\in\mathbb{R}$, ${\bm\beta}_M^*\in\mathbb{R}^p$, and ${\bm\beta}_I^*\in\mathbb{R}^{n\times[p(p-1)/2]}$ defined analogously. Let ${\mathcal M}^*=\{j: \beta_j^*\not=0\}$ and ${\mathcal I}^*=\{(j,k):\beta_{jk}^*\not=0,1\le j<k\le p\}$ be the sets of the important main effects and interactions, respectively. Then, the true HDGLM is
\begin{equation}
\label{eq:true high-dimensional GLM with interaction effects}
g({\bm\mu}^*)={\bm\eta}^*={\bf X}{\bm\beta}^*=\beta_0^*+\sum_{j\in{\mathcal M}^*} {\bm x}_j\beta_j^*+\mathop{\sum\sum}_{(j,k)\in{\mathcal I}^*} {\bm x}_j\circ{\bm x}_k\beta_{jk}^*.
\end{equation}
The true model is unknown. One can only assume that $|{\mathcal M}^*|$ and $|{\mathcal I}^*|$ are small with $|{\mathcal M}^*|+|{\mathcal I}^*|\ll n$ in the true model. A research task is to select the important main effects and interactions from~\eqref{eq:high-dimensional GLM with interaction effects}, such that the selected model can correctly reflect~\eqref{eq:true high-dimensional GLM with interaction effects} with a high probability. 

For selection of interactions from an ultrahigh-dimensional statistical model, it is important to consider well-established restrictions, such as heredity~\cite{chipman1996,hamada1992} and marginality~\cite{mccullagh2002} interpreted by the {\it strong hierarchy} (also called strong heredity) restriction. It requires that the selected model should satisfy $\hat\beta_{jk}\not=0\Rightarrow \hat\beta_j\hat\beta_k\not=0$ even when the SH restriction is violated in the true model, where $\hat\beta_j$ and $\hat\beta_{jk}$ are the estimates of $\beta_j$ and $\beta_{jk}$ provided by a variable selection method, respectively. Violations of hierarchy restrictions can only occur in special situations, implying that they should be treated as the default~\cite{bien2013}. A few previous methods have been developed based on this requirement. Examples include the strong heredity interaction model (\textsf{SHIM})~\cite{choi2010}, the variable selection using adaptive nonlinear interaction structures in high dimensions (\textsf{VANISH})~\cite{radchenko2010}, the hierarchical net (\textsf{hierNet})~\cite{bien2013}, the group regularized estimation under structural hierarchy (\textsf{GRESH})~\cite{she2018}, the high-dimensional quadratic regression (\textsf{HiQR})~\cite{wang2024}, the sparse reluctant interaction modeling (\textsf{sprinter})~\cite{yu2023}, the \textsf{hierScale}~\cite{hazimeh2020}, the regularization algorithm under the marginality principle (\textsf{RAMP})~\cite{hao2018}, and the stepwise conditional likelihood variable selection for discriminant analysis (\textsf{SODA})~\cite{li2019}. We study those and find that the SH restriction may be violated in the \textsf{HiQR}, the \textsf{sprinter}, and the \textsf{SODA}, implying that they are inappropriate. It is unlikely to use the \textsf{SHIM}, the \textsf{hierNet}, the \textsf{RAMP}, and the \textsf{hierScale} to identify important interactions if the associated main effects are weak or absent. Although it is addressed by the \textsf{GRESH} and the \textsf{VANISH}, they may include too many unimportant main effects or interactions in their selected models. We propose the SH $L_0$ (i.e., \textsf{SHL0}) method to address this issue. We show that the \textsf{SHL0} satisfies the SH restriction, enjoys the oracle properties, and is computationally efficient. The details are displayed in Section~\ref{sec:method}.

In the literature, variable selection using the $L_0$ penalty has only been studied for the main effects. It has been pointed out that using the $L_0$ penalty is more essential for variable selection than the well-known $L_1$~\cite{tibshirani1996} and the nonconcave~\cite{fanli2001,zhang2010} penalties. The main issue is that the $L_0$ penalty is discontinuous at $0$, leading to difficulty in optimizing the corresponding objective function by a coordinate descent algorithm. Previous methods seek approximations that closely resemble the $L_0$ penalized objective function. The goal is to formulate a coordinate descent algorithm to optimize the objective function approximately. Examples include the mixed integer optimization~\cite{bertsimas2016}, the seamless $L_0$~\cite{dicker2013}, the $L_0$EM~\cite{liu2016}, the APM$L_0$~\cite{li2018}, and the subderivative~\cite{huang2018,kim2021}. The derivation of the approximations involves both the likelihood function and the $L_0$ penalty. They cannot be used to select interactions if the response does not follow normal. We devise a local combinatorial optimization (LCO) algorithm to overcome the difficulty. An advantage is that the LCO can be applied to arbitrary ultrahigh-dimensional statistical models with interactions. 

Combinatorial optimization is one of the most important in theoretical computer science. It has been widely applied in many fields, including artificial intelligence, machine learning, and software engineering. The goal is to find an optimal object from a finite set of candidates. Combinatorial optimization has played an important role in discovering a well-known concept called NP-completeness. The discovery was based on a study on the {\it Boolean satisfiability problem}~\cite{cook1971,levin1973}. Before the study,  the concept of NP-completeness even did not exist. Many other NP-complete combinatorial optimization problems have been found. Examples include the {\it assignment problem}, the {\it integer programming problem}, the {\it knapsack problem}, the {\it traveling salesman problem}, and the {\it minimum spanning tree problem}; to name a few. It is impossible to solve these problems by polynomial algorithms except one can prove one side of the well-known open problem ${\rm P}={\rm NP}$, while the other side is ${\rm P}\not={\rm NP}$. To address the computational challenges in real-world applications, combinational optimization is usually implemented by a local search called LCO~\cite{crama2005,orlin2004}. A concern is that the local search can only provide a locally optimal solution. To increase the possibility of a global solution, a straightforward approach is to carry out an algorithm multiple times, implying that LCO is usually nondeterministic. The LCO can well-address the computational challenges. 

The article is organized as follows. In Section~\ref{sec:related work}, we review related work. In Section~\ref{sec:method}, we introduce our method. In Section~\ref{sec:theoretical properties}, we evaluate the theoretical properties of our method. In Section~\ref{sec:simulation}, we compare our method with our competitors via Monte Carlo simulations. In Section~\ref{sec:application}, we apply our method to a real data example. In Section \ref{sec:conclusion}, we conclude. We put all of the proofs in the Appendix.

\section{Related Work}
\label{sec:related work}

The earliest work for variable selection of interactions under the SH restriction was the \textsf{SHIM}~\cite{choi2010}. For an ultrahigh-dimensional linear model (HDLM) as
\begin{equation}
\label{eq:full model with interaction effects}
{\bm y}={\bf X}{\bm\beta}+{\bm\epsilon}=\beta_0+\sum_{j=1}^p {\bm x}_j\beta_j+\sum_{j=1}^{p-1}\sum_{k=j+1}^p {\bm x}_j\circ{\bm x}_k\beta_{jk}+{\bm\epsilon}, {\bm\epsilon}\sim {\cal N}({\bm 0},\phi{\bf I}),
\end{equation}
the \textsf{SHIM} selects the main effects and interactions by minimizing the loss
\begin{equation}
\label{eq:objective function of the shin}
{\cal L}_{\bm\lambda}^{SHIM}({\bm\beta})={1\over 2}\|{\bm y}-{\bf X}{\bm\beta}\|^2+n\lambda_2\sum_{j=1}^p |\beta_j|+n\lambda_1\sum_{j=1}^{p-1}\sum_{k=j+1}^p |\gamma_{jk}|,
\end{equation}
where $\gamma_{jk}$ is derived by reparametrizing $\beta_{jk}=\gamma_{jk}\beta_j\beta_k$, and ${\bm\lambda}=(\lambda_1,\lambda_2)^\top$ with $\lambda_1,\lambda_2>0$ as two tuning parameters. The goal of reparametrization is to ensure that the SH restriction is satisfied. Optimization of the \textsf{SHM} is carried out by a coordinate descent algorithm. As pointed out by the authors, the coordinate descent algorithm may be stuck in a local optimizer because ${\cal L}_{\bm\lambda}^{SHIM}({\bm\beta})$ is not concave. In the iterations, the \textsf{SHIM} encounters a problem of $\hat\beta_j\hat\beta_k=0\Rightarrow \hat\gamma_{jk}=0$, leading to difficulty in selecting an important interaction if the associated main effects are weak or absent. The coordinate descent algorithm is applied to the set of all variables. The computation is time-consuming when $p\gg n$ in~\eqref{eq:objective function of the shin}. The difficulty can be overcome by incorporating the \textsf{AcorSIS}~\cite{zhang2024}.

Both  the \textsf{hierNet}~\cite{bien2013} and the \textsf{GRESH}~\cite{she2018} and are modified from the \textsf{VANISH}~\cite{radchenko2010}. The \textsf{hierNet} enforces $\sum_{k=1}^p |\beta_{jk}|\le |\beta_j|$ for all $j\in\{1,\dots,p\}$ to satisfy the SH restriction. Similar to the \textsf{SHIM}, the computation of the \textsf{hierNet} proposed by the authors is time-consuming when $p\gg n$. The problem can be solved by incorporating the \textsf{AcorSIS}. Based on some $q\in[1,\infty]$, the \textsf{GRESH} selects the main effects and interactions from~\eqref{eq:full model with interaction effects} by minimizing
\begin{equation}
\label{eq:objective function of GRESH}
{\mathcal L}_{{\bm\lambda}q}^{GRESH}({\bm\beta})={1\over 2}\|{\bm y}-{\bf X}{\bm\beta}\|^2+n\lambda_2\sum_{j=1}^p \|{\bm\beta}_j\|_q+n\lambda_1\sum_{j=1}^{p-1}\sum_{k=j+1}^p |\beta_{jk}|,
\end{equation}
where ${\bm\beta}_j=(\beta_j,\beta_{j1},\dots,\beta_{j(j-1)},\beta_{j(k+1)},\cdots\beta_{jp})^\top$. The second term on the right-hand side of~\eqref{eq:objective function of GRESH} is proposed for the main effects. It ensures that $\beta_j$ cannot be penalized to be zero if there is $\beta_{jk}\not=0~\exists k\not=j$. The objective function given by~\eqref{eq:objective function of GRESH} is defined for arbitrary $q\in[1,\infty]$, but the corresponding alternative direction method of multiplier (ADMM) algorithm is proposed for $q=2$. A concern is that the ADMM has a constant unrelated to the objective function of the \textsf{GRESH}. The coordinate descent algorithm proposed by \cite{zhang2024} can address the concern. Direct optimization of ${\mathcal L}_{{\bm\lambda}q}^{GRESH}({\bm\beta})$ is time-consuming when $p\gg n$. The problem can be addressed by incorporating the \textsf{AcorSIS}.

The objective function of the \textsf{hierScale}~\cite{hazimeh2020} is equivalent to ${\mathcal L}_{{\bm\lambda}\infty}^{GRESH}({\bm\beta})$. The computation for the \textsf{hierScale} proposed by the authors does not suffer from the computational difficulties that appeared in the \textsf{SHIM} and the \textsf{GRESH} when $p\gg n$ because of two stages: screening and penalization. The screening stage eliminates redundant variables. The penalization stage selects the main effects and interactions from a shrunk set of active variables derived by the screening stage. Still, it is hard to detect important interactions if one of its related main effects is weak or absent. This is overcome by the proposed \textsf{SHL0}. Similar issues also occur in the \textsf{RAMP}~\cite{hao2018}. The \textsf{RAMP} is computationally efficient because of the two stages for the main effects and interactions respectively. Starting from an empty set of main effects and interactions, the \textsf{RAMP} adds main effects first and interactions next. It requires that interactions be associated with the main effects. Thus, the \textsf{RAMP} satisfies the SH restriction.

Both the \textsf{HiQR}~\cite{wang2024} and the \textsf{sprinter}~\cite{yu2023} are developed under a framework called the all-pair LASSO (APL). 
The APL selects the main effects and interactions from~\eqref{eq:full model with interaction effects} by minimizing 
\begin{equation}
\label{eq:all pair LASSO}
{\mathcal L}_{\bm\lambda}^{APL}({\bm\beta})={1\over 2}\|{\bm y}-{\bf X}{\bm\beta}\|^2+n\lambda_2 \sum_{j=1}^p |\beta_j|+n\lambda_1\sum_{j=1}^{p-1}\sum_{k=j+1}^p |\beta_{jk}|.
\end{equation}
To improve the computational efficiency, the \textsf{HiQR} and the \textsf{sprinter} impose all-pair SIS~\cite{hallxue2014} to eliminate redundant effects before the implementation of the APL. Because the interactions are ignored when the main effects are penalized, a violation of the SH restriction may be induced. We investigated this issue in our simulations. We identified this concern (see Section~\ref{sec:simulation} for the details).

The \textsf{SODA}~\cite{li2019} is a stepwise method for selecting the main effects and interactions from an ultrahigh-dimensional logistic model (HDLOGIT) for ${\bm y}\sim{\mathcal Bin}({\bm m},{\bm\pi})$, ${\bm m}=(m_1,\dots,m_n)^\top\in\mathbb{N}^n$ and ${\bm\pi}=(\pi_1,\dots,\pi_n)^\top\in(0,1)^n$, with 
\begin{equation}
\label{eq:ultrahigh-dimensional logistic model}
\log{{\bm\pi}\over {\bm 1}-{\bm\pi}}={\bf X}{\bm\beta}=\beta_0+\sum_{j=1}^p {\bm x}_j\beta_j+\sum_{j=1}^{p-1}\sum_{k=p+1}^p {\bm x}_j\circ{\bm x}_k\beta_{jk}.
\end{equation}
A concern is that the \textsf{SODA} may violate the SH restriction. We find this in our numerical study (see Section~\ref{sec:simulation} for the details).

\section{Method}
\label{sec:method}

The associated main effects of the interaction ${\bm x}_j\circ{\bm x}_k$ are ${\bm x}_j$ and ${\bm x}_k$. The associated interactions of the main effect ${\bm x}_j$ are ${\bm x}_j\circ{\bm x}_k$ for all $k\not=j$. We say that ${\bm x}_j$ is an important main effect if $j\in{\mathcal M}^*$ or ${\bm x}_j\circ{\bm x}_k$ is an important interaction if $(j,k)\in{\mathcal I}^*$. We say that ${\bm x}_j$ is associated with an important interaction if $(j,k)\in{\mathcal I}^*$ or $(k,j)\in{\mathcal I}^*$ for some $k\not=j$. We say ${\bm x}_j$ is an active variable if ${\bm x}_j$ is an important main effect or associated with an important interaction. If ${\bm x}_j$ is an inactive variable, then ${\bm x}_j$ is an unimportant main effect and unrelated to any important interaction either. 

We use $\alpha$ to represent a reduced HDGLM derived by eliminating many main effects and interactions from~\eqref{eq:high-dimensional GLM with interaction effects}. Then, $\alpha$ is composed of a subset of the main effects as $\alpha_M\subset\{1,\cdots,p\}$ and a subset of the interactions as $\alpha_I\subseteq\{(j,k): 1\le j<k\le p\}$, implying that it can be expressed as $\alpha=\alpha_M\cup\alpha_I$.  The strong hierarchy is satisfied if $(j,k)\in\alpha$ only when $j\in\alpha$ and $k\in\alpha$. As the strong hierarchy may not be satisfied in~\eqref{eq:true high-dimensional GLM with interaction effects}, we modify the true HDGLM as $\alpha^*=\alpha_M^*\cup\alpha_I^*$ with $\alpha_I^*={\mathcal I}^*$ and $\alpha_M^*=\{j: j\in{\mathcal M}^*\ {\rm or}\ \exists\ k\not=j\ {\rm such\ that}\ (j,k)\in{\mathcal I}^*\ {\rm or}\ (k,j)\in{\mathcal I}^* \}$. The goal is to ensure that $\alpha^*$ satisfies the strong hierarchy. Thus, the associated main effects of $\alpha_I^*$ form a subset of $\alpha_M^*$.  We use $\alpha^*$ to evaluate the theoretical properties. 

\subsection{SHPML}
\label{subsec:SHPML}

The main effects are selected by the penalized maximum likelihood (PML), which becomes the penalized least squares (PLS) if the response is normal. The PML is proposed when the third term on the right-hand side of~\eqref{eq:high-dimensional GLM with interaction effects} is absent. The objective function of the PML is $\ell_\lambda(\beta_0,{\bm\beta}_M;\phi)=\ell(\beta_0,{\bm\beta}_M;\phi)-(n/2)\sum_{j=1}^p P_\lambda(|\beta_j|)$,
where $\ell(\beta_0,{\bm\beta}_M;\phi)={\bf 1}^\top\{{\bm y}\circ h({\bm\eta})/\phi-b[h({\bm\eta})]/\phi+c({\bm y},\phi)\}$ is the loglikelihood function of the HDGLM jointly defined by~\eqref{eq:distribution of exponential family} and~\eqref{eq:high-dimensional GLM with interaction effects} by assuming $\beta_{jk}=0$ for all $j\not=k$, ${\bm\theta}=h({\bm\eta})$ is the inverse function of $g[b'({\bm\theta})]={\bm\eta}$, $P_\lambda(\cdot)$ is a penalty function, and $\lambda> 0$ is a tuning parameter. Various functions can be used to specify $P_\lambda(\cdot)$. Examples include the LASSO~\cite{tibshirani1996}, the SCAD~\cite{fanli2001}, and the MCP~\cite{zhang2010}. They are popular in selecting the main effects from an ultrahigh-dimensional statistical model. 

Due to the requirement of the SH restriction, variable selection methods for the interactions cannot be straightforwardly extended from the main effects. We devise the SH penalized maximum likelihood (SHPML) method for variable selection of interactions with the SH restriction. We specify it for the \textsf{SHL0} method using the $L_0$ penalty.

The SHPML is proposed for the case when $p\gg n$. It is composed of two stages: screening and penalization. We cannot use the previous \textsf{AcorSIS} to screen variables when the response does not follow normal. We investigate two possible extensions of the \textsf{AcorSIS}. The first, called the aggregated likelihood ratio sure independence screening (\textsf{ALRSIS}), is devised based on likelihood ratio statistics. The aggregated score sure independence screening (\textsf{ASSIS}), which is the second, is proposed based on score statistics. The \textsf{ASSIS} is more computationally efficient than the \textsf{ALRSIS}. 

The previous \textsf{AcorSIS} can be classified as a marginal screening method because it is developed based on a marginal utility. According to~\cite{barutfan2016}, marginal screening methods can have large false positives, indicating that they may recruit many variables with strong marginal utilities but conditionally independent of the response given other variables. To address this issue, conditional screening is recommended. Following this line, we propose both  the \textsf{ALRSIS} and the \textsf{ASSIS} methods under a base model $\alpha$ satisfying the SH restriction as
\begin{equation}
\label{eq:initial model for SIS}
g({\bm\mu}_\alpha)={\bm\eta}_{\alpha}={\bf X}_\alpha{\bm\beta}_\alpha=\beta_0+\sum_{j\in\alpha_{M}}{\bm x}_j\beta_j+\mathop{\sum\sum}_{(j,k)\in\alpha_{I},j<k}{\bm x}_j\circ{\bm x}_k \beta_{jk},
\end{equation}
where ${\bf X}_\alpha$ with the corresponding regression coefficient vector ${\bm\beta}_\alpha$ is the sub-matrix derived by eliminating ${\bm x}_j$ for all $j\not\in\alpha_{\mathcal M}$ and ${\bm x}_{j}\circ{\bm x}_k$ for all $(j,k)\not\in\alpha_{\mathcal I}$ from ${\bf X}$. The goal is to catch important ${\bm x}_j$ outside of $\alpha_{M}$. To achieve the research goal, for each $j\not\in\alpha_M$, we individually compare the divergence between the base model and a set of expanded models.  For any distinct $j,k\in\bar\alpha_M\cup\{0\}$ with $\bar\alpha_M$ representing the complementary subset of $\alpha_M$, we define an expanded model for ${\bm x}_j\circ{\bm x}_k$ with ${\bm x}_j={\bm x}_0\circ{\bm x}_j$ as 
\begin{equation}
\label{eq:expanded model interaction j and k}
g({\bm\mu}_{\alpha\cup\{(j,k)\}})={\bm\eta}_{\alpha\cup\{(j,k)\}}={\bf X}_\alpha{\bm\beta}_\alpha+{\bm x}_j\circ{\bm x}_k\beta_{jk},
\end{equation}
where $k<j$ is allowed. We use either the likelihood ratio or the score statistics to compare the aggregated divergence between the base model and the expanded models, leading to the \textsf{ALRSIS} and the \textsf{ASSIS} methods for the screening stage,  respectively. Both the \textsf{ALRSIS} and the \textsf{ASSIS} are conditional screening methods. If we choose $\alpha=\emptyset$, then they become marginal screening methods. 

The development of the \textsf{ALRSIS} method is straightforward. The likelihood statistics between the base and the expanded models are available in many software packages. An example is the \textsf{glm} function of \textsf{R}. Let $G_{\alpha}^2$ and  $G_{\alpha\cup\{(j,k)\}}^2$ be the residual deviance values of~\eqref{eq:initial model for SIS} and \eqref{eq:expanded model interaction j and k}, respectively. Then,  $G_\alpha^2-G_{\alpha\cup\{(j,k)\}}^2$ is the likelihood ratio statistic for divergence between \eqref{eq:initial model for SIS} and \eqref{eq:expanded model interaction j and k}. The aggregated likelihood ratio statistic for ${\bm x}_j$ with $j\in\bar\alpha_M$ is
\begin{equation}
\label{eq:aggregated likelihood ratio}
aG_{j|\alpha}^2=\max_{k\in\bar\alpha_M\cup\{0\},k\not=j}\{ G_\alpha^2-G_{\alpha\cup\{(j,k)\}}^2\}.
\end{equation}
We expect $aG_{j|\alpha}^2$ to be large if ${\bm x}_j$ is an active variable outside of $\alpha_M$; or small otherwise. We define a likelihood ratio shrunk  subset of variables as
\begin{equation}
\label{eq:shrunk subset of variable likelihood ratio}
{\mathcal A}_{\alpha}^{LR}=\{j\not\in\alpha_M: aG_{j|\alpha}^2~{\rm is~among~the~first~}d_\gamma~{\rm largest~of~all}\},
\end{equation}
where $d_\gamma=[{\gamma}n]$ denotes the integer part of ${\gamma}n$. The convenient value is $\gamma=1/\log{n}$, as it has been widely used previously~\cite[e.g.]{kongli2017}. 

We next introduce our \textsf{ASSIS} method. The idea is to replace the likelihood ratio statistic with the score statistic for the divergence between~\eqref{eq:initial model for SIS} and \eqref{eq:expanded model interaction j and k}. In particular, denote $\ell({\bm\beta}_{\alpha\cup\{(j,k)\}};\phi)$ as the loglikelihood function of~\eqref{eq:expanded model interaction j and k}. Let $\dot\ell({\bm\beta}_{\alpha\cup\{(j,k)\}};\phi)$ be the corresponding gradient vector with respect to ${\bm\beta}_{\alpha\cup\{(j,k)\}}$ and $I({\bm\beta}_{\alpha\cup\{(j,k)\}};\phi)$ be the Fisher Information. The score statistic for divergence between  \eqref{eq:initial model for SIS} and \eqref{eq:expanded model interaction j and k} is
\begin{equation}
\label{eq:score statistics between two models}
S_{(j,k)|\alpha}=\dot\ell^\top(\breve{\bm\beta}_{\alpha\cup\{(j,k)\}};\hat\phi_\alpha)I^{-1}(\breve{\bm\beta}_{\alpha\cup\{(j,k)\}};\hat\phi_\alpha)\dot\ell(\breve{\bm\beta}_{\alpha\cup\{(j,k)\}};\hat\phi_\alpha),
\end{equation}
where $\breve{\bm\beta}_{\alpha\cup\{(j,k)\}}=(\hat{\bm\beta}_\alpha^\top,0)^\top$ and $\hat{\bm\beta}_\alpha$ and $\hat\phi_\alpha$ are the MLEs of ${\bm\beta}_\alpha$ and $\phi$ under~\eqref{eq:initial model for SIS}. The aggregated score statistic for ${\bm x}_j$ with $j\in\bar\alpha_M$ is 
\begin{equation}
\label{eq:aggregated score statistic}
aS_{j|\alpha}=\max_{k\in\bar\alpha_M\cup\{0\},k\not=j}\{ S_{(j,k)|\alpha}\}.
\end{equation}
Similarly, we expect that $aS_{j|\alpha}$ is large if ${\bm x}_j$ is an active variable outside of $\alpha_M$; or small otherwise, leading to a score shrunk subset of variables as
\begin{equation}
\label{eq:shrunk subset of variable score}
{\mathcal A}_{\alpha}^{S}=\{j\not\in\alpha_M: aS_{j|\alpha}~{\rm is~among~the~first~}d_\gamma~{\rm largest~of~all}\},
\end{equation}
where $d_\gamma$ is the same as~\eqref{eq:shrunk subset of variable likelihood ratio}.

\begin{thm}
\label{thm:formulation for the aggregated score statistic}
Let ${\bm r}_\alpha= ({\bf I}-{\bf P}_\alpha){\bf W}_\alpha^{1/2}({\bm z}_\alpha-{\bf X}_\alpha{\bm\beta}_\alpha)  $ and ${\bm s}_{(j,k)|\alpha}=({\bf I}-{\bf P}_\alpha){\bf W}_\alpha^{1/2}{\bm x}_{j}\circ{\bm x}_{k}$, where ${\bm z}_\alpha=\hat{\bm\eta}_\alpha+({\bm y}-\hat{\bm\mu}_\alpha)\circ{\partial\hat{\bm\eta}_\alpha}/\partial\hat{\bm\mu}_{\alpha} $, ${\bf W}_\alpha=\phi{\rm diag}\{(\partial\hat{\bm\mu}_\alpha/\partial\hat{\bm\eta}_\alpha )^2/b''(\hat{\bm\eta}_\alpha)\}$, $\hat{\bm\mu}_\alpha=b'(\hat{\bm\theta}_\alpha) $, $\hat{\bm\theta}_\alpha$ and $\hat{\bm\eta}_\alpha$ satisfy $g[b'(\hat{\bm\theta}_\alpha)]=\hat{\bm\eta}_\alpha={\bf X}_\alpha\hat{\bm\beta}_\alpha$, and ${\bf P}_\alpha={\bf W}_\alpha^{1/2}{\bf X}_\alpha({\bf X}_\alpha^\top {\bf W}_\alpha{\bf X}_\alpha)^{-1} {\bf X}_\alpha^\top{\bf W}_\alpha^{1/2}$. Then, 
\begin{equation}
\label{eq:another expression of the aggregated score statistic}
\eqalign{
aS_{j|\alpha}
=&\max_{j,k\in\bar\alpha_M\cup\{0\},k\not=j} {\hat\phi_\alpha \langle  {\bm r}_\alpha,{\bm s}_{(j,k)|\alpha} \rangle \over \|{\bm s}_{(j,k)|\alpha}\|}.
}
\end{equation}
\end{thm}

We compare the computational properties of the \textsf{ALRSIS} with the \textsf{ASSIS} methods. The likelihood ratio statistic $G_\alpha^2-G_{\alpha\cup\{(j,k)\}}^2$ involves the MLEs of ${\bm\beta}_\alpha$ and ${\bm\beta}_{\alpha\cup\{(j,k)\}}$. The derivation involves two models. The computation of $aG_j^2$ for ${\bm x}_j$ given by~\eqref{eq:aggregated likelihood ratio} involves the MLEs of $p-|\alpha_M|$ models. The computation of ${\mathcal A}_\alpha^{LR}$ given by~\eqref{eq:shrunk subset of variable likelihood ratio} involves the MLEs of $(p-|\alpha_M|)(p-|\alpha_M|-1)/2$ models. If the \textsf{ALRSIS} method is adopted, then the MLEs of $O(p^2)$ models are needed. Note that $aS_{j|\alpha}$ given by~\eqref{eq:another expression of the aggregated score statistic} only involves the MLE of ${\bm\beta}_\alpha$. It is enough for ${\mathcal A}_\alpha^S$ given by~\eqref{eq:shrunk subset of variable score}. If the \textsf{ASSIS} is adopted, then only the MLE of ${\bm\beta}_\alpha$ is needed. The property significantly affects the computational efficiencies of the two methods. 
We numerically studied this issue based on the HDLOGIT given by~\eqref{eq:ultrahigh-dimensional logistic model} with $n=200$. The study showed that the time taken were $2.13$, $8.88$, $54.37$, $06.39$, and $843.65$ minutes when $p=100,200,500,1000,2000$ respectively in the \textsf{ALRSIS} method. The time taken reduced to $4\times10^{-4},1.6\times 10^{-3}, 0.01,0.04,0.18$ minutes in the \textsf{ASSIS} method. Therefore, we recommend the \textsf{ASSIS} method in practice. 

We present the penalization stage of the \textsf{SHL0} method with the \textsf{ASSIS} is adopted in the screening stage. We use $\|{\bm\xi}_j\|_{0,mod}=I({\bm\xi}_j\not={\bm 0})$ with ${\bm\xi}_j=(\beta_j,\beta_{1j},\cdots,\beta_{(j-1)j},\beta_{(j+1)j},\dots,\beta_{pj})^\top$ representing a parameter vector composed of main effects and interactions associated to ${\bm x}_j$ for $j=1,\dots,p$ and $\|{\bm t}\|_{0,mod}$ is the modified $L_0$-norm (i.e., the discrete norm) defined in a vector space as $\|{\bm t}\|_{0,mod}=0$ if ${\bm t}={\bm 0}$ or $\|{\bm t}\|_{0,mod}=1$ otherwise to control the main effects. The modified $L_0$-norm satisfies $\|{\bm\xi}_j\|_{0,mod}=1$ iff $\beta_j\not=0$ or there is at least one nonzero $\beta_{jk}$ or $\beta_{kj}$, $k\not=j$. The penalization stage likely preserves $\beta_j$ if ${\bm x}_j$ is associated with at least one of the important interactions. The interactions can be naturally controlled by $\|\beta_{jk}\|_0=I(\beta_{jk}\not=0)$ for any distinct $j$ and $k$. We propose the objective function of the \textsf{SHL0} in the penalization stage as
\begin{equation}
\label{eq:objective function L0 interaction effect}
\eqalign{
\ell_\lambda({\bm\beta};\phi)=&\ell({\bm\beta};\phi)-{n\lambda\over 2}\sum_{j\in\mathcal A} \|{\bm\xi}_j\|_{0,mod}-{n\lambda\over 2}\mathop{\sum\sum}_{j,k\in{\mathcal A},j<k} \|\beta_{jk}\|_0,\cr
}
\end{equation}
where we choose ${\mathcal A}=\alpha_M\cup{\mathcal A}_{\alpha}^{LR}$ for the \textsf{ALRSIS} method and ${\mathcal A}=\alpha_M\cup{\mathcal A}_{\alpha}^{S}$ for the \textsf{ASSIS} method in the screening stage.  The default is ${\mathcal A}=\alpha_M\cup{\mathcal A}_{\alpha}^{S}$.

Based on a given $\lambda$, we estimate the parameters by
\begin{equation}
\label{eq:estimator of the L0 PML}
\{\hat{\bm\beta}_\lambda,\hat\phi_\lambda\}=\mathop{\arg\!\max}_{{\bm\beta},\phi}\ell_\lambda({\bm\beta};\phi),
\end{equation}
leading to the estimate of the model as $\hat\alpha_\lambda=\hat\alpha_{M\lambda}\cup\hat\alpha_{I\lambda}$ with $\hat\alpha_{M\lambda}=\{j: \hat\beta_{j\lambda}\not=0\}$ and $\hat\alpha_{I\lambda}=\{(j,k): \hat\beta_{jk\lambda}\not=0\}$, where $\hat\beta_{j\lambda}$ is the estimate of $\beta_j$ and $\hat\beta_{jk\lambda}$ is the estimate of $\beta_{jk}$ given by~\eqref{eq:estimator of the L0 PML}. We use those to formulate the estimates of $\beta_{0}$, ${\bm\beta}_{M}$, and ${\bm\beta}_{I}$, denoted by $\hat\beta_{0\lambda}$, $\hat{\bm\beta}_{M\lambda}$, and $\hat{\bm\beta}_{I\lambda}$, respectively. If $\phi$ is not present in~\eqref{eq:distribution of exponential family}, then we can simply remove $\hat\phi_\lambda$ and $\phi$ form the left-hand and the right-hand sides of~\eqref{eq:estimator of the L0 PML}, respectively, implying that~\eqref{eq:estimator of the L0 PML} can also be applied if ${\bm y}$ is a binomial or a Poisson response vector. 

The intercept term is not penalized. The main effects are penalized by the second term on the right-hand side of~\eqref{eq:objective function L0 interaction effect}. For each $j$,  it likely penalizes $\beta_j$ to be $0$ if the main effect ${\bm x}_j$ and the interactions ${\bm x}_j\circ{\bm x}_k$ for all $k\not=j$ are weak. If one of $\beta_{jk}$ is strong, there is likely to have $\|{\bm\xi}_j\|_{0,mod}=1$, leading to $\beta_j$ contained in the selected model.  We assume that the explanatory variables are collected from continuous distributions to prove this property theoretically. We treat~\eqref{eq:high-dimensional GLM with interaction effects} as conditional given ${\bm x}_1,\dots,{\bm x}_p$. 

\begin{thm}
\label{thm:strong hierarchy satisfied}
If ${\bm x}_1,\dots,{\bm x}_p$ are continuous random vectors and $\lambda>0$, then $\hat\alpha_\lambda$ given by~\eqref{eq:estimator of the L0 PML} satisfies the SH restriction with probability one, and $\hat{\bm\beta}_\lambda$ and $\hat\phi_\lambda$ are identical to the MLEs of ${\bm\beta}$ and $\phi$ under $\hat\alpha_\lambda$ with probability one. 
\end{thm}

The tuning parameter is unknown. It should be determined before the application of~\eqref{eq:objective function L0 interaction effect} and~\eqref{eq:estimator of the L0 PML}. We use $\lambda$ to control the main effects and interactions simultaneously. The previous methods use two distinct tuning parameters to control those separately. It is considered one of the advantages of our method over the previous methods. 
 
\subsection{Tuning Parameter}
\label{subsec:tuning parameter}

In the literature, the determination of tuning parameters is investigated by the asymptotic theory. The goal is to ensure that the resulting estimators are consistent and the variable selection method enjoys oracle properties~\cite{fanli2001}. The generalized information criterion (GIC)~\cite{zhangli2010} is the most appropriate approach for selecting $\lambda$ in~\eqref{eq:objective function L0 interaction effect}. It is carried out by minimizing the GIC function as
\begin{equation}
\label{eq:GIC objective function}
{\rm GIC}_{\kappa}(\lambda)=-2\ell(\hat{\bm\beta}_\lambda;\hat\phi_\lambda)+\kappa{df}_\lambda
\end{equation}
with the best choice of $\lambda$ given by
\begin{equation}
\label{eq:the best lambda in L0 interaction}
\hat\lambda=\mathop{\arg\!\min}_\lambda {\rm GIC}_{\kappa}(\lambda),
\end{equation}
where $df_{\lambda}$ is the model degrees of freedom defined as the number of nonzero components contained by $\hat{\bm\beta}_{\lambda}$ and $\kappa$ is a pre-selected constant that controls the properties of the GIC. If~\eqref{eq:the best lambda in L0 interaction} is used, the solutions of~\eqref{eq:estimator of the L0 PML} are denoted as $\hat{\bm\beta}_{\hat\lambda}$, $\hat\phi_{\hat\lambda}$, and $\hat\alpha_{\hat\lambda}$, respectively. Large values of $\kappa$ induce lower values of $df_{\hat\lambda}$ in the GIC, implying that an appropriate choice of $\kappa$ is needed. 

The selection of $\kappa$ in the GIC has been well-studied for the case when interactions are absent. Examples include the AIC by $\kappa=2$ and the BIC by $\kappa=\log{n}$ for the low-dimensional setting, where it assumes that $p$ is bounded as $n\rightarrow\infty$. The high-dimensional setting assumes that $p$ grows with $n$. For the cases when $p$ polynomially and exponentially increases with $n$ as $n\rightarrow\infty$, the HBIC~\cite{wangkim2013} by $\kappa=2\log{p}$ or the EBIC~\cite{fantang2013} by $\kappa=\log{p}\log\!\log{n}$ are recommended. Due to the presence of the interactions, the problem should be re-investigated. We summarize our findings by the following theorem.

\begin{thm}
\label{thm:consistency of L0 penalty}
Suppose that all assumptions of Theorem~\ref{thm:goodness of the fit statistic of the true model} of Section~\ref{sec:theoretical properties} hold. Assume that one of (i) $p$ is fixed, (ii) $p=O(n^d)$ for some $d>0$, or (iii) $p=O(e^{n^d})$ for some $d<1/2$ is satisfied as $n\rightarrow\infty$. If $\kappa=\log{n}$ in (i) or $\kappa=(4+\epsilon)\log(n\vee p)$ for some $\epsilon>0$ or $\kappa=\log{p}\log\!\log{n}$ in (ii) or (iii), then $\hat\alpha_{\hat\lambda}\stackrel{P}\rightarrow\alpha^*$ and $\hat{\bm\beta}_{\hat\lambda}\stackrel{P}\rightarrow{\bm\beta}^*$.
\end{thm}

Although our interest is the high-dimensional setting, we still consider the low-dimensional setting in Theorem~\ref{thm:consistency of L0 penalty}. If $p$ is fixed as $n\rightarrow\infty$, then we recommend the BIC by $\kappa=\log{n}$. In the high-dimensional setting, we assume that $p>n$ with $\kappa=4\log{p}$ or $\kappa=\log{p}\log\!\log{n}$. We show that~\eqref{eq:the best lambda in L0 interaction} can be analytically solved if $\kappa$ is provided. The base is the relationship between the \textsf{SHL0} and the GIC objective functions as
\begin{equation}
\label{eq:connection between the L0 penalized and GIC functions}
{\rm GIC}_{\kappa}({\kappa\over n})=-2\ell_{\kappa/ n}(\hat{\bm\beta}_{\kappa/n};\hat\phi_{\kappa/n}).
\end{equation} 
It induces the following theorems.

\begin{lem}
\label{lem:monotonicity of GIC}
There exist $0=\lambda_{v}< \lambda_{v-1}<\cdots <\lambda_0<\lambda_{-1}= \infty$ for some $v\le p\wedge n$ almost surely such that ${\rm GIC}_\kappa(\lambda)$ strictly decreases in but $|\hat\alpha_\lambda|$ does not vary with $\lambda\in[\lambda_{s-1},\lambda_s)$ for each $s=0,1,\dots, v$. If $\lambda$ increases from a value lower than $\lambda_s$ to a value higher than that, then the reduction of $|\hat\alpha_\lambda|$ is greater than or equal to $1$.
\end{lem}

\begin{thm}
\label{thm:solution of lambda}
$\hat\lambda=\mathop{\arg\!\min}_{\kappa/n\in[\lambda_s,\lambda_{s-1})}\lambda$. 
\end{thm}

Theorem~\ref{thm:solution of lambda} points out that $\hat\lambda=\lambda_s$ if $[\lambda_s,\lambda_{s-1})$ contains $\kappa/n$. The shrinkage from $\kappa/n$ to $\lambda_s$ changes ${\rm GIC}_{\kappa}(\lambda)$ but not the loglikelihood value, leading to the following theorem. 

\begin{cor}
\label{cor:GIC and L0 penalty}
If $\hat\lambda$ is solved by~\eqref{eq:the best lambda in L0 interaction}, then $\hat{\bm\beta}_{\kappa/n}=\hat{\bm\beta}_{\hat\lambda}$ and $\hat\phi_{\kappa/n}=\hat\phi_{\hat\lambda}$, and $\hat\lambda$ is the minimum of $\lambda$ which induces  $\hat{\bm\beta}_{k/n}$ and $\hat\phi_{\kappa/n}$ as the final answer of~\eqref{eq:estimator of the L0 PML}.
\end{cor}

Corollary~\ref{cor:GIC and L0 penalty} means that we can simply use $\lambda=[\log{n}\vee(4\log{p})]/n$ or $\lambda=(1/n)\log{p}\log\!\log{n}$ in the implementation of~\eqref{eq:objective function L0 interaction effect} and~\eqref{eq:the best lambda in L0 interaction}, implying that the tuning parameter has a closed form expression. This cannot be achieved if other penalties are used. Examples include the \textsf{glmnet} and \textsf{ncvreg} packages of \textsf{R}, which usually contain $100$ options of $\lambda$ in their output. The \textsf{SHL0} has an obvious advantage in selecting tuning parameters. 

\subsection{Computation}
\label{subsec:computation}

In the penalization stage, the main issue is the computation of $\hat\alpha_\lambda$ for a given $\lambda$, because $\hat{\bm\beta}_\lambda$ and $\hat\phi_\lambda$ are identical to the MLEs of ${\bm\beta}$ and $\phi$ under $\hat\alpha_\lambda$.  The $L_0$ penalty is not continuous. The usual coordinate descent algorithm cannot be applied. To address this issue, previous work approximates the $L_0$ penalized objective function~\cite{huang2018}. In this research, we discard the coordinate descent algorithm and propose a local combinatorial optimization (LCO) algorithm.

Combinatorial optimization~\cite{du1998} is fundamental in theoretical computer science, but it has not been applied in variable selection problems. The computational of global solutions of many well-known combinatorial optimization problems is usually NP-hard~\cite{papadimitriou1998}. It is more appropriate to devise a nondeterministic approach to address the challenge, leading to the proposed LCO. The goal of nondeterministic is to increase the probability of deriving a global solution by replications. We show that our algorithm is polynomial and enjoys nice theoretical properties.

We want to derive the best $\alpha$ which maximizes $\ell_\lambda(\hat{\bm\beta}_\alpha;\hat\phi_\alpha)$ by replications. The computation of the best $\alpha$ can be treated as one of the combinatorial optimization algorithms because the number of possible values of  $\ell_\lambda(\hat{\bm\beta}_\alpha;\hat\phi_\alpha)$ is lower than $2^{|{\mathcal A}|({\mathcal A}|+1)/2}$. A global solution may be derived via a brute-force algorithm for all the possibilities. This is unrealistic if $|{\mathcal A}|$ is not extremely small. An LCO algorithm is recommended.

The LCO starts with an initial $\alpha$ and the corresponding $\hat{\bm\beta}_\alpha$ and $\hat\phi_\alpha$. It searches the best $\tilde\alpha$ in the neighborhood of $\alpha$ to update $\alpha$. In particular, let 
\begin{equation}
\label{eq:definition of neighbor of SH restriction}
{\mathcal N}(\alpha)=\{\tilde\alpha\in\mathscr{A}:{\rm there~is~no}~\check\alpha\in\mathscr{A}~{\rm~such~that}~\alpha\subset\check\alpha\subset\tilde\alpha~{\rm or}~ \tilde\alpha\subset\check\alpha\subset\alpha\}
\end{equation}
be the neighbor of $\alpha$,  where $\mathscr{A}$ represents the collection of all models that satisfy the SH restriction. We use the difference operator 
\begin{equation}
\label{eq:difference operation}
{\rm diff}(\alpha,\tilde\alpha)=\left\{\begin{array}{ll} \{j\}, & j\in {\mathcal A}_M,\cr \{(j,k)\}, &  (j,k)\in{\mathcal A}_I, \cr \end{array}\right.
\end{equation}
 to describe the relationship between $\alpha\in\mathscr{A}$ and $\tilde\alpha\in{\mathcal N}(\alpha)$, where ${\mathcal A}_M=\{j:j\in{\mathcal A}\}$ and  ${\mathcal A}_I=\{(j,k):j,k\in{\mathcal A},j<k\}$. There are two scenarios in~\eqref{eq:difference operation}. In the first scenario, there is $\tilde\alpha\subset\alpha$. Then, $\tilde\alpha$ is derived by eliminating a main effect or an interaction from $\alpha$. We only remove $(j,k)\in\alpha_I$ from $\alpha$ if the interaction ${\bm x}_j\circ{\bm x}_k$ is chosen, leading to ${\rm diff}(\alpha,\tilde\alpha)=\{(j,k)\}$. In this case, we have $\tilde\alpha_M=\alpha_M$ and $\tilde\alpha_{I}=\alpha_I\backslash\{(j,k)\}$ for some $(j,k)\in\alpha_I$. We remove $j\in\alpha_M$ and all $(j,k)\in\alpha_I$ when $j<k$ or $(k,j)\in\alpha_I$ when $k<j$ from $\alpha$ if the main effect ${\bm x}_j$ is chosen, leading to ${\rm diff}(\alpha,\tilde\alpha)=\{j\}$. In this case, we have $\tilde\alpha_M=\alpha_M\backslash\{j\}$ and $\tilde\alpha_I=\alpha_I\backslash(\{(j,k): (j,k)\in\alpha\}\cup \{(k,j): (k,j)\in\alpha\})$ for some $j\in\alpha_M$. In the second scenario, there is $\alpha\subset\tilde\alpha$. Then, $\tilde\alpha$ is derived by adding a main effect or an interaction. We add the corresponding $j\not\in\alpha_M$ to $\alpha$ if the main effect ${\bm x}_j$ is chosen, leading to ${\rm diff}(\alpha,\tilde\alpha)=\{j\}$. In this case, we have $\tilde\alpha_M=\alpha_M\cup\{j\}$ and $\tilde\alpha_I=\alpha_I$ for some $j\not\in\alpha_M$. We add $j,k$ and $(j,k)\not\in\alpha_I$ to $\alpha$ if the interaction ${\bm x}_j\circ{\bm x}_m$ is chosen, leading to ${\rm diff}(\alpha,\tilde\alpha)=\{(j,k)\}$. In this case, we have $\tilde\alpha_M=\alpha_M\cup\{j,k\}$ and $\tilde\alpha_I=\alpha_I\cup\{(j,k)\}$ for some $(j,k)\not\in\alpha_I$. 
Therefore, ${\rm diff}(\alpha,\tilde\alpha)$ given by~\eqref{eq:difference operation} is well-defined, and $\tilde\alpha\in{\mathcal N}(\alpha)$ is uniquely determined by ${\rm diff}(\alpha,\tilde\alpha)$ for any $j\in{\mathcal A}_M$ or $(j,k)\in{\mathcal A}_I$.

To decide whether we should update $\alpha$ with some $\tilde\alpha\in{\mathcal N}(\alpha)$, we use the penalized likelihood ratio statistic as
\begin{equation}
\label{eq:penalized likelihood ratio statistic for main effects}
\Lambda_\lambda(\tilde\alpha)=\ell_{\lambda}(\hat{\bm\beta}_{\tilde\alpha};\hat\phi_{\tilde\alpha})-\ell_{\lambda}(\hat{\bm\beta}_\alpha;\hat\phi_\alpha)
\end{equation}
to measure the difference between $\alpha$ and $\tilde\alpha$, which prefers $\tilde\alpha$ if $\Lambda_\lambda(\tilde\alpha)>0$. Because~\eqref{eq:penalized likelihood ratio statistic for main effects} should be examined for all $\tilde\alpha\in{\mathcal N}(\alpha)$, we search the best $\tilde\alpha$. This is related to the local search approach in the literature of theoretical computer science~\cite{crama2005}. There are two strategies. The first is the greedy strategy called the best one local search (B1LS) strategy. It uses 
\begin{equation}
\label{eq:best one local search}
\tilde\alpha_{B1}=\mathop{\arg\!\max}_{\tilde\alpha\in{\mathcal N}(\alpha)}\Lambda_\lambda(\tilde\alpha)
\end{equation}
to update $\alpha$. The second is the first one local search (F1LS) strategy, which uses
\begin{equation}
\label{eq:first one local search}
\tilde\alpha_{F1}=\mathop{\arg\!\min}_{\tilde\alpha\in{\mathcal N}(\alpha)}\{\tau(\tilde\alpha):\Lambda_\lambda(\tilde\alpha)>0 \}
\end{equation}
where $\tau(\tilde\alpha)$ is an order of $\tilde\alpha$ in ${\mathcal N}(\alpha)$. Any order $\tau(\tilde\alpha)$ can be used in~\eqref{eq:first one local search}. An example is the natural order defined as $\tilde\alpha'\prec\tilde\alpha''$ if ${\rm diff}(\alpha,\tilde\alpha')\prec{\rm diff}(\alpha,\tilde\alpha'')$, where we define $(j,k)\prec(j',k')$ if $j<j'$ or $j=j'$ and $k<k'$. In this research, we discard the natural order and adopt the random order for the nondeterministic in the LCO. A random permutation on the natural order derives a random order. The stopping rules of the B1LS and the F1LS strategies are identical. We recommend the F1LS strategy because it is more computationally efficient.

We devise an algorithm with both the screening and penalization stages. We adopt the \textsf{ASSIS} method in the screening stage and the F1LS strategy in the penalization stage. We screen the variables with an initial $\alpha$ and the default $\alpha=\emptyset$, leading to ${\mathcal A}_\alpha^S$ by~\eqref{eq:shrunk subset of variable score}. We derive ${\mathcal A}=\alpha_M\cup {\mathcal A}_\alpha^S$ at the end of the screening stage. We reinitialize $\alpha$ with $\alpha=\emptyset$ at the beginning of the penalization stage. We apply $\ell_\lambda({\bm\beta};\phi)$ given by~\eqref{eq:objective function L0 interaction effect} to the F1LS strategy with a random order of $\tau(\tilde\alpha)$ on ${\mathcal N}(\alpha)$. To enhance the probability of deriving a global solution, we repeat the application of the F1LB strategy a few times (the default is $10$). We choose the one with the greatest $\ell_\lambda({\bm\beta};\phi)$ value as the output of the penalization stage. To reduce the risk of missing important variables in the screening stage, we use the output of the penalization stage to set the initial $\alpha$ and redo the screening and penalization stages again. We obtain Algorithm~\ref{alg:two stage algorithm for interaction effects}.

\begin{algorithm}
\caption{\label{alg:two stage algorithm for interaction effects} \textsf{SHL0} for selecting the main effects and and interactions}
\begin{algorithmic}[1]
\Statex{{\bf Input}: ${\bm y}\in\mathbb{R}^n$, $({\bm x}_1,\dots,{\bm x}_p)\in\mathbb{R}^{n\times p}$, and $\lambda$}
\Statex{{\bf Output}:  $\hat{\bm\beta}_\lambda$, $\hat\phi_\lambda$, and $\hat\alpha_\lambda$}
\State{Initialize $\alpha$ (the default is $\alpha=\emptyset$)}
\State{Screening stage: derive ${\mathcal A}_{\alpha}^S$ by~\eqref{eq:shrunk subset of variable score} and set ${\mathcal A}=\alpha_M\cup {\mathcal A}_{\alpha}^S$}
\Statex{Penalization stage: }
\State{Reinitialize $\alpha$ (the default is $\alpha=\emptyset$)}
\State{Generate a random order on ${\mathcal A}_M\cup{\mathcal A}_I$}
\State{Compute $\tilde\alpha$ sequentially based on the random order, and update $\alpha\leftarrow\tilde\alpha$ if  $\Lambda_\lambda(\tilde\alpha)>0$ until all elements in ${\mathcal A}_M\cup{\mathcal A}_I$ are examine}
\State{Repeat Steps 4 and 5 until $\alpha$ cannot be updated, and then stop the penalization stage}
\State{The second round: initialize $\alpha$ by the output of Step 6 and redo Steps 2-6}
\State {Output}
\end{algorithmic}
\end{algorithm}

The screening stage is not necessary when $p$ is small. It is critical when $p$ is large, especially when $p\gg n$. We cannot guarantee the final solution of Algorithm~\ref{alg:two stage algorithm for interaction effects} is a global optimizer. We devise Step 4 to incorporate the nondeterministic, which is a common approach in theoretical computer science. We devise Step 7 to capture important variables that may be missed in the screening stage of the first round of computation. It can be extended for an algorithm with more than two rounds. Our research shows that the second round may be needed but the third round is usually unnecessary. Therefore, we only redo the screening and penalization stages one more times, leading to Step 7 in our algorithm.


\subsection{Extension}
\label{subsec:extension}

Only maximum likelihood algorithms and model degrees of freedom are needed. Our method can be straightforwardly extended to arbitrary ultrahigh-dimensional statistical models. We assume that the PDF or PMF of $y_i$ can be expressed as $f(y_i;\eta_i,\phi)$ with $\eta_i$ is modeled by the right-hand side of~\eqref{eq:high-dimensional GLM with interaction effects}. If the MLEs $\hat{\bm\beta}_\alpha$ and $\hat\phi_\alpha$ of ${\bm\beta}$ and $\phi$ can be calculated based on a given $\alpha$, then we can modify Algorithm~\ref{alg:two stage algorithm for interaction effects} for variable selection with interactions to determine the best $\alpha$. This is considered in our theoretical studies.

\section{Theoretical Properties}
\label{sec:theoretical properties}

Our main interest is the high-dimensional setting, where $p$ grows with $n$ as $n\rightarrow\infty$. We still study the low-dimensional setting, where $p$ is bounded as $n\rightarrow\infty$. We show that our method enjoys the oracle properties if we choose $\lambda=[\log{n}\vee(4\log{p})]/n$ or $\lambda=(1/n)\log{p}\log\!\log{n}$ in~\eqref{eq:estimator of the L0 PML}. We assumed that the screening stage is needed only in the high-dimensional setting.  We focus on the case when $\phi$ is present in~\eqref{eq:distribution of exponential family} because it implies the case when $\phi$ is absent. Only theoretical properties of the maximum likelihood are needed. We extend our conclusions to general ultrahigh-dimensional statistical models introduced in Section~\ref{subsec:extension}. 

We use $\Omega$ to denote the set of all candidate models. We assume that there is an upper bound $K=o(n)$ for the number of variables a candidate model contains, implying that we have $|\alpha|\le K$ for all $\alpha\in\Omega$. The true model is unique in $\Omega$, meaning that $\alpha\in\Omega$ is the true model iff $\alpha=\alpha^*$. We say $\alpha$ is underfitted if $\alpha\not\supset\alpha^*$ or overfitted if $\alpha\supset\alpha^*$.  We partition $\Omega$ into the set of underfitted models denoted by $\Omega_{-}=\{\alpha: \alpha\not\supset\alpha^*\}$, the set of overfitted models denoted by $\Omega_{+}=\{\alpha:\alpha\supset\alpha^*\}$, and the set of the true model denoted by $\Omega_*=\{\alpha:\alpha=\alpha^*\}$. Thus, $|\Omega_*|=1$. 

To simply our notations, we use ${\bm\beta}_\alpha$ to represent ${\bm\beta}$ restricted on $\alpha$, meaning that ${\bm\beta}_\alpha$ is derived by setting $\beta_j=0$ for $j\not\in\alpha$ and $\beta_{jk}=0$ for $(j,k)\not\in\alpha$. The dimension of ${\bm\beta}_\alpha$ is still $1+p(p+1)/2$. The MLEs of ${\bm\beta}_\alpha$ and $\phi$ under $\alpha$, denoted by $\hat{\bm\beta}_\alpha$ and $\hat\phi_\alpha$ respectively, are restricted on the main effects and interactions in $\alpha$, implying that the dimension of $\hat{\bm\beta}_\alpha$ is $|\alpha|+1$. 

We assume that $g(\cdot)$ is the canonical link, such that the loglikelihood function of~\eqref{eq:high-dimensional GLM with interaction effects} under $\alpha$ can be expressed as
\begin{equation}
\label{eq:loglikelihood under alpha parameter}
\eqalign{
\ell({\bm\beta}_\alpha;\phi)={\bf 1}^\top\left[{{\bm y}\circ{\bm\eta}_{\alpha}-b({\bm\eta}_\alpha)\over\phi}+c({\bm y},\phi)\right],\cr
}\end{equation}
where  ${\bm\mu}_{\alpha}=g^{-1}({\bm\eta}_\alpha)$, ${\bm\eta}_{\alpha}={\bf X}{\bm\beta}_\alpha$, and ${\bf X}\in\mathbb{R}^{n\times[1+p(p+1)/2]}$ is the design matrix of~\eqref{eq:high-dimensional GLM with interaction effects}. The score function of $\ell({\bm\beta}_\alpha;\phi)$ with respect to ${\bm\beta}_\alpha$ is $\dot\ell({\bm\beta}_\alpha;\phi)={\bf X}_{\alpha}^\top({\bm y}-{\bm\mu}_\alpha)/\phi$, where ${\bf X}_\alpha$ is a submatrix of ${\bf X}$ after removing columns for $j\not\in\alpha_M$ and $(j,k)\not\in\alpha_I$. The Hessian matrix of  $\ell({\bm\beta}_\alpha;\phi)$ with respect to ${\bm\beta}_\alpha$, denoted by $-{\cal I}_n({\bm\beta}_\alpha)/\phi$, is the negative Fisher Information matrix times by $n$. It satisfies
\begin{equation}
\label{eq:hessian of loglikelihood}
-n{\cal I}_n({\bm\beta}_\alpha)=\phi\ddot\ell({\bm\beta}_\alpha;\phi)=-{\bf X}_\alpha^\top{\rm diag}\left({\partial{\bm\mu}_{\alpha}\over\partial{\bm\eta}_\alpha}\right){\bf X}_\alpha=-{\bf X}_\alpha^\top{\rm diag}[b''({\bm\eta}_\alpha)]{\bf X}_\alpha.
\end{equation}
Because ${\bm x}_j$ is standardized and the right-hand side of~\eqref{eq:hessian of loglikelihood} does not depend on the response,  we assume that ${\cal I}({\bm\beta}_\alpha)=\lim_{n\rightarrow\infty}{\cal I}_n({\bm\beta}_\alpha)$ exists for any $\alpha\in\Omega$.  It is enough to display our proofs based on~\eqref{eq:loglikelihood under alpha parameter}, implying that presenting those for other links is unnecessary.

We need a functional central limit theorem (FCLT) to prove our conclusions. We study it for overfitted models. Let ${\bm\beta}_\alpha^*$ be a sub-vector of ${\bm\beta}^*$ after $\beta_j$ for all $j\not\in\alpha$ and $\beta_{jk}$ for all $(j,k)\not\in\alpha_I$ are removed. Then, the dimension of $\hat{\bm\beta}_\alpha$ and ${\bm\beta}_\alpha^*$ are all equal to $|\alpha|+1$. If $\alpha\in\Omega_+\cup\Omega_*$, then ${\rm E}[\dot\ell({\bm\beta}_\alpha^*)]={\bm 0}$. Using the Taylor approximation, we have
\begin{equation}
\label{eq:taylor expansion MLE at true}
\ell(\hat{\bm\beta}_{\alpha};\phi)=\ell({\bm\beta}_\alpha^*;\phi)+{1\over 2\phi}n(\hat{\bm\beta}_\alpha-{\bm{\bm\beta}_\alpha^*})^\top{\cal I}({\bm\beta}_\alpha^*)(\hat{\bm\beta}_\alpha-{\bm{\bm\beta}_\alpha^*})+o_p(1).
\end{equation}
  If ${\cal I}^{-1}({\bm\beta}_\alpha^*)$ exists, then under suitable conditions, we have $\sqrt{n/\phi^*}(\hat{\bm\beta}_\alpha-{\bm{\bm\beta}_\alpha^*})\rightsquigarrow {\cal N}({\bf 0},{\cal I}_\alpha^{-1})$. Using the properties of the likelihood ratio statistic~\cite[P. 145]{ferguson1996}, we have
\begin{equation}
\label{eq:asymptotic distribution of the likelihood ratio given alpha}
\eqalign{
\Lambda_\alpha=2[\ell(\hat{\bm\beta}_{\alpha};\phi^*)-\ell(\hat{\bm\beta}_{\alpha^*};\phi^*)]=&{n\over\phi^*}(\hat{\bm\beta}_\alpha-{\hat{\bm{\bm\beta}}_{\alpha^*}^\alpha})^\top{\cal I}_\alpha(\hat{\bm\beta}_\alpha-{\hat{\bm{\bm\beta}}_{\alpha^*}^\alpha})+o_p(1)
\rightsquigarrow \chi_{|\alpha|-|\alpha^*|}^2
}
\end{equation}
and  
\begin{equation}
\label{eq:asymptotic distribution of the MLE in the selected and true model}
\sqrt{n\over\phi^*}(\hat{\bm\beta}_\alpha-\hat{\bm\beta}_{\alpha^*}^\alpha)\rightsquigarrow {\cal N}({\bf 0},{\bm\Sigma}_\alpha),
\end{equation}
where ${\cal I}_\alpha={\cal I}({\bm\beta}_\alpha^*)$, ${\bm\Sigma}_\alpha={\cal I}_\alpha^{-1}({\bf I}-{\cal I}_\alpha{\cal Q}_\alpha ){\cal I}_\alpha ({\bf I}-{\cal I}_\alpha{\cal Q}_\alpha ) {\cal I}_\alpha^{-1}$, $\hat{\bm\beta}_{\alpha^*}^\alpha$ is the $(|\alpha|+1)$-dimensional vector expanded from the $(|\alpha^*|+1)$-dimensional vector $\hat{\bm\beta}_{\alpha^*}$ (i.e., the MLE of ${\bm\beta}$ under $\alpha^*$) by adding zeros, and ${\cal Q}_\alpha$ is the $(|\alpha|+1)\times(|\alpha|+1)$-dimensional matrix expanded from the $(|\alpha^*|+1)\times(|\alpha^*|+1)$-dimensional matrix ${\cal I}_{\alpha^*}^{-1}$ by adding zeros. It is appropriate to use $\Lambda_\alpha$ to test whether $\alpha$ can be reduced to $\alpha^*$ for any overfitted $\alpha$. 

As $\alpha\in\Omega_+$ can be arbitrary in the test, we evaluate the joint limiting distribution of $\sqrt{n}(\hat{\bm\beta}_{\alpha_1}-\hat{\bm\beta}_{\alpha^*}^{\alpha_1})$ and $\sqrt{n}(\hat{\bm\beta}_{\alpha_2}-\hat{\bm\beta}_{\alpha^*}^{\alpha_2})$ for all distinct $\alpha_1,\alpha_2\in\Omega_+$. If ${\cal I}({\bm\beta}_{\alpha_1},{\bm\beta}_{\alpha_2})=\lim_{n\rightarrow\infty}{\cal I}_n({\bm\beta}_{\alpha_1},{\bm\beta}_{\alpha_2})$ exists, where ${\cal I}_n({\bm\beta}_{\alpha_1},{\bm\beta}_{\alpha_2})= n^{-1}{\bf X}_{\alpha_1}^\top{\rm diag}\{[b''({\bm\eta}_{\alpha_1})]^{1/2}\}{\rm diag}\{[b''({\bm\eta}_{\alpha_2})]^{1/2}\}{\bf X}_{\alpha_2}$, then under suitable regularity conditions we have
\begin{equation}
\label{eq:joint distribution of MLE of two models}
\sqrt{n\over\phi^*}\left[\left(\begin{array}{c}\hat{\bm\beta}_{\alpha_1}\cr  \hat{\bm\beta}_{\alpha_2}\cr \end{array}  \right)- \left(\begin{array}{c} \hat{\bm\beta}_{\alpha^*}^{\alpha_1} \cr \hat{\bm\beta}_{\alpha^*}^{\alpha_2} \cr \end{array}   \right)  \right]\rightsquigarrow {\cal N}\left[\left(\begin{array}{c}{\bf 0}\cr{\bf 0} \end{array}   \right),  \left(\begin{array}{cc} {\bm\Sigma}_{\alpha_1}
 &   {\bm\Sigma}_{\alpha_1\alpha_2} \cr  {\bm\Sigma}_{\alpha_2\alpha_1} & {\bm\Sigma}_{\alpha_2} \cr   \end{array}\right)  \right],
\end{equation}
where ${\bm\Sigma}_{\alpha_1\alpha_2}={\cal I}_{\alpha_1}^{-1}({\bf I}-{\cal I}_{\alpha_1}{\cal Q}_{\alpha_1} ){\cal I}_{\alpha_1\alpha_2} ({\bf I}-{\cal I}_{\alpha_2}{\cal Q}_{\alpha_2} ) {\cal I}_{\alpha_2}^{-1}$ and ${\cal I}_{\alpha_1\alpha_2}={\cal I}({\bm\beta}_{\alpha_1}^*,{\bm\beta}_{\alpha_2}^*)$. All of the asymptotic distributions given by the above can be simultaneously obtained by the FCLT of the partial sum given below.

\begin{lem}
\label{lem:basic FCLT without the linear transformation}
(FCLT of the partial sum). Let $S_{njk}=({\bm x}_j\circ{\bm x}_k)^\top({\bm y}-{\bm\mu}^*)=\sum_{i=1}^n x_{ij}x_{ik}(y_i-\mu_i^*)$ for $j,k=0,\dots,p$, where ${\bm x}_0={\bf 1}$ is the $n$-dimensional vector with all components equal to $1$. If $\lim_{n\rightarrow\infty}\sum_{i=1}^n (x_{ij}x_{ik})^2b''(\theta_{i}^*)/n=\sigma_{jk}^2>0$ exists and ${\rm E}(|x_{ij}x_{ik}(y_i-\mu_i^*)|^3)$ is uniformly bounded, then $S_{[nt]jk}/(\sigma_{jk} n^{1/2})\rightsquigarrow \mathbb{W}_{jk}$ for each $j,k=0,\dots,p$, where $\mathbb{W}_{jk}$ is the standard Brownian motion on $[0,1]$.
\end{lem}

\begin{thm}
\label{thm:asymptotic distributions of the terms}
If the assumptions of Lemma~\ref{lem:basic FCLT without the linear transformation} are satisfied, then~\eqref{eq:asymptotic distribution of the MLE in the selected and true model} holds for any $\alpha\in\Omega_+\cup\Omega_*$,~\eqref{eq:asymptotic distribution of the likelihood ratio given alpha} holds for any $\alpha\in\Omega_+$,  and~\eqref{eq:joint distribution of MLE of two models} holds for any distinct $\alpha_1,\alpha_2\in\Omega_+\cup\Omega_*$.
\end{thm}

\begin{lem}
\label{lem:asymptotic distribution of the likelihood ratio statistics many models}
Suppose that all of the assumptions of Lemma~\ref{lem:basic FCLT without the linear transformation} hold. Assume that ${\cal I}^{-1}({\bm\beta}_\alpha)$ exists and is invertible for all $\alpha\in\Omega_{+}\cup\Omega_*$ and ${\cal I}({\bm\beta}_{\alpha_1},{\bm\beta}_{\alpha_2})$ exists for all $\alpha_1,\alpha_2\in\Omega_{+}\cup\Omega_*$. Then, there exist unique mean zero normal random vectors ${\bm z}_{\alpha_1},\dots,{\bm z}_{\alpha_m}$ satisfying ${\rm V}({\bm z}_{\alpha_i})={\bm\Sigma}_{\alpha_i}$ and ${\rm cov}({\bm z}_{\alpha_i},{\bm z}_{\alpha_j})={\bm\Sigma}_{\alpha_i\alpha_j}$ for any distinct $\alpha_1,\dots,\alpha_m\in\Omega_{+}\cup \Omega_*$, where $m$ is a finite positive integer.
\end{lem}

\begin{cor}
\label{cor:joint limit distribution of likelihood ratio}
If the assumptions of Lemma~\ref{lem:asymptotic distribution of the likelihood ratio statistics many models} are satisfied, then $(\Lambda_{\alpha_1},\dots,\Lambda_{\alpha_m})\rightsquigarrow(q_{\alpha_1},\dots,q_{\alpha_m})$ for any distinct $\alpha_1,\cdots\alpha_m\in\Omega_{+}$, where $q_{\alpha_i}={\bm z}_{\alpha_i}^\top{\cal I}_{\alpha_i}{\bm z}_{\alpha_i}\sim\chi_{|\alpha_i|-|\alpha*|}^2$ for all $i=1,\dots,m$.
\end{cor} 

We use the FCLT to evaluate the impact of underfitted models. We expand an underfitted model to an overfitted model by adding missing main effects and interactions, such that the FCLT  overfitted models can be used. In particular, for any underfitted $\alpha$, we expand it to overfitted $\tilde\alpha=\alpha\cup\alpha^*$ and evaluate the Taylor expansion of $\ell({\bm\beta}_{\alpha};\phi)$ at ${\bm\beta}^*$ as
\begin{equation}
\label{eq:taylor expansion at the true}
\eqalign{
\ell({\bm\beta}_{\alpha};\phi)
=&\ell({\bm\beta}_{\alpha}^*;\phi)+{1\over\phi}T_1({\bm\beta}_{\alpha})-{1\over 2\phi}T_2({\bm\beta}_{\alpha})+{1\over2\phi}R({\bm\beta}_{\alpha},\check{\bm\beta}_{\tilde\alpha}),\cr
}\end{equation}
where $T_1({\bm\beta}_{\alpha})=[{\bf X}_{\tilde\alpha}({\bm\beta}_{\alpha}-{\bm\beta}^*)]^\top({\bm y}-{\bm\mu}^*)$, $T_2({\bm\beta}_{\alpha})=n({\bm\beta}_{\alpha}-{\bm\beta}^*)^\top{\cal I}_n({\bm\beta}^*)({\bm\beta}_{\alpha}-{\bm\beta}^*)$, $R({\bm\beta}_{\alpha},\check{\bm\beta}_{\alpha})=-(1/6)\sum_{i=1}^n b'''(\check\eta_{i\tilde\alpha})(\eta_{i\alpha}-\eta_{i}^*)^3$, $\check{\bm\mu}_{\tilde\alpha}=g^{-1}(\check{\bm\eta}_{\tilde\alpha})$, $\check{\bm\eta}_{\tilde\alpha}=(\check\eta_{1\tilde\alpha},\dots,\check\eta_{n\tilde\alpha})^\top={\bf X}_{\tilde\alpha}\check{\bm\beta}_{\alpha}$, and $\check{\bm\beta}_{\alpha}$ is between ${\bm\beta}^*$ and ${\bm\beta}_{\alpha}$ studied in $\tilde\alpha$. For the second term on the right-hand side of~\eqref{eq:taylor expansion at the true}, we have ${\rm E}_{{\bm\beta}^*}[T_1({\bm\beta}_{\alpha})]={\rm E}_{{\bm\beta}^*}\{[{\bf X}({\bm\beta}_{\alpha}-{\bm\beta}^*)]^\top({\bm y}-{\bm\mu}^*)\}=0$ and ${\rm V}_{{\bm\beta}^*}[T_1({\bm\beta}_{\alpha})]={\rm V}_{{\bm\beta}^*}\{[{\bf X}({\bm\beta}_{\alpha}-{\bm\beta}^*)]^\top({\bm y}-{\bm\mu}^*)\}=T_2({\bm\beta}_{\alpha})$, leading to the following Lemma.

\begin{lem}
\label{lem:the linear term}
If all of the eigenvalues of ${\cal I}({\bm\beta}_{\tilde\alpha})$ are between $c_1$ and $c_2$ for some positive $c_1$ and $c_2$, then $T_1({\bm\beta}_{\alpha})=o_p(T_2({\bm\beta}_{\alpha},\check{\bm\beta}_{\tilde\alpha}))$ for any sequence of ${\bm\beta}_{\alpha}$ satisfying $\lim\!\inf_{n\rightarrow\infty}n^{1/2-\epsilon}\|{\bm\beta}_{\alpha}-{\bm\beta}^*\|>0$, where $T_2({\bm\beta}_{\alpha},\check{\bm\beta}_{\tilde\alpha})=-(1/2)T_2({\bm\beta}_{\alpha})+T_3({\bm\beta}_{\alpha},\check{\bm\beta}_{\tilde\alpha})$ and $\epsilon$ can be any real number in $(0,1/2)$. 
\end{lem}

Lemma~\ref{lem:the linear term} indicates that the second term on the right-hand side of~\eqref{eq:taylor expansion at the true} can be ignored in the evaluation of the Taylor expansion. Because $\alpha$ is underfitted, ${\bm\beta}^*$ is outside of the domain of ${\bm\beta}_{\alpha}$, implying that the minimum value of $T_2({\bm\beta}_{\alpha})$ can be large as $n\rightarrow\infty$. The magnitude is evaluated by the Kullback-Leibler (KL) divergence at ${\bm\beta}_\alpha$ as ${\rm KL}({\bm\beta}_\alpha)=\phi{\rm E}[\ell({\bm\beta}^*;\phi)-\ell({\bm\beta}_{\alpha};\phi)]=(1/2)T_2({\bm\beta}_{\alpha})$, which can be used for both overfitted and underfitted $\alpha$. To ensure identifiability, we assume that ${\rm KL}({\bm\beta}_\alpha)$ has a unique minimizer, denoted by ${\bm\psi}_\alpha$, for every $\alpha$ satisfying $|\alpha|\le K$. We use $\tilde{\bm\psi}_\alpha\in\mathbb{R}^{1+p(p+1)/2}$ to represent the vector expanded from ${\bm\psi}_\alpha$ by adding $0$ for columns not contained by $\alpha$. If $\alpha$ is overfitted, then $\tilde\alpha=\alpha$, $\min_{{\bm\beta}_\alpha}[{\rm KL}({\bm\beta}_\alpha)]=0$, and $\tilde{\bm\psi}_\alpha={\bm\beta}^*$. If $\alpha$ is underfitted, then $\tilde{\bm\psi}_{\bm\alpha}\not={\bm\beta}^*$ and $\min_{{\bm\beta}_\alpha}{\rm KL}({\bm\beta}_\alpha)>0$, implying that ${\rm KL}({\bm\beta}_\alpha)$ measures the deviance from the true model $\alpha^*$ due to missing at least one main effect $\beta_j^*$ or interaction effect $\beta_{jk}^*$. We use
\begin{equation}
\label{eq:KL for underfitted}
\delta_n
=\inf_{\alpha_1\cap\alpha^*\subset\alpha_2\cap\alpha^*}{{\rm KL}({\bm\beta}_{\alpha_1})-{\rm KL}({\bm\beta}_{\alpha_2})\over n|(\alpha_2\cap\alpha^*)\backslash(\alpha_1\cap\alpha^*)|}
\end{equation}
to measure the smallest signal strength from ${\bm\beta}^*$. It effectively controls the magnitude of the difference of the likelihood ratio statistics between the true $\alpha^*$ and the entire set of underfitted models. A similar approach has been previously used by~\cite{fantang2013,zhangli2010}. To implement~\eqref{eq:KL for underfitted}, we choose $K=p(p+1)/2$ in the low-dimensional setting or $K=o(n)$ as $n\rightarrow\infty$ in the high-dimensional setting.

\begin{thm}
\label{thm:consistency under the low-dimensional setting}
(Consistency and oracle properties under the low-dimensional setting). Suppose all assumptions of Lemma~\ref{lem:asymptotic distribution of the likelihood ratio statistics many models} are satisfied and the condition of Lemma~\ref{lem:the linear term} is also satisfied. If $p$ is fixed as $n\rightarrow\infty$, $\mathop{\lim\!\inf}_{n\rightarrow\infty} n^{v}\delta_n>0$ for some $v<1/2$, and $\lambda=n^{-1}\log{n}$, then $\hat\alpha_\lambda\stackrel{P}\rightarrow \alpha^*$ and~\eqref{eq:estimator of the L0 PML} satisfies the oracle properties.
\end{thm}

We next study the high-dimensional setting. We assume that the \textsf{SHL0} method contains the screening stage. For the screening stage, the main issue is to show $\lim_{n\rightarrow\infty} {\rm Pr}(\alpha_M^*\subseteq{\cal A}_\alpha^{LR})=1$ and $\lim_{n\rightarrow\infty} {\rm Pr}(\alpha_M^*\subseteq{\cal A}_\alpha^S)=1$ under a few suitable regularity conditions (i.e., consistency and also called sure screening property). For the penalization stage, the main issue is to show $\hat\alpha\stackrel{P}\rightarrow\alpha^*$ as $n\rightarrow\infty$ (also called consistency). We study the two problems separately because the application of the \textsf{ALRSIS} and the \textsf{ASSIS} is not limited to the penalization stage of the \textsf{SHL0}. 

Because $|\Omega|\rightarrow\infty$ as $n\rightarrow\infty$, we cannot use the method in the proof of Theorem~\ref{thm:consistency under the low-dimensional setting}. To address this issue, we need to the FCLT. Because ${\bm z}_\alpha$ can be treated as by a linear transformation on $\mathbb{W}$, the FCLT holds in the entire $\Omega_+\cup\Omega_*$. If $\alpha\in\Omega_-$, then we expand it to an overfitted $\tilde\alpha$. By the Kolmogorov existence theorem~\cite{billingsley1995}, we can use the covariance function ${\bm\Sigma}_{\alpha_1\alpha_2}$ to construct a unique mean zero Gaussian random field $\mathbb{G}$, such that there is ${\rm cov}[ \mathbb{G}(\alpha_1),\mathbb{G}(\alpha_2)]={\bm\Sigma}_{\alpha_1\alpha_2}$ for any distinct $\alpha_1,\alpha_2$. If the FCLT holds, then we have $\sqrt{n/\phi^*}(\hat{\bm\beta}_\alpha-{\bm\psi}_\alpha)\rightsquigarrow\mathbb{G}(\alpha)$ uniformly in the entire $\Omega_+$. To show the FCLT, we need $\sqrt{n/\phi^*}(\hat{\bm\beta}_{\alpha_1}-\hat{\bm\psi}_{\alpha_1},\dots,\hat{\bm\beta}_{\alpha_k}-\hat{\bm\psi}_{\alpha_k})\rightsquigarrow(\mathbb{G}(\alpha_1),\dots,\mathbb{G}(\alpha_k))$ for any finite $k$ and the uniform convergence of $\sqrt{\phi^*/n}{\bf X}_\alpha^\top[\dot\ell({\bm\beta}_\alpha)-{\bm\psi}_{\alpha}]$ for all $\alpha\in\Omega$. The uniform convergence is obtained by the maximal inequality~\cite[P. 284]{vandervaart1998}, which is the major issue in the proof of the FCLT. If the FCLT holds, then for all $\alpha\in\Omega$, the maximum of the absolute difference between $\sqrt{\phi^*/n}{\bf X}_\alpha^\top[\dot\ell({\bm\beta}_\alpha)-{\bm\psi}_\alpha]$ and its limiting distribution uniformly converges to $0$ in probability. 

\begin{lem}
\label{lem:FCLT under high-dimensional setting}
Suppose that all assumptions of Lemma~\ref{lem:asymptotic distribution of the likelihood ratio statistics many models} are satisfied, the condition of Lemma~\ref{lem:the linear term} is also satisfied, and $\sum_{i=1}^n {\rm E}(|x_{ij}x_{ik}(y_i-\mu_i^*)|^3)/\sum_{i=1}^n (x_{ij}x_{ik})^2b''(\theta_i^*)$ for any $j,k=0,1,\dots,p$ is uniformly bounded. Then, $\sqrt{n/\phi^*}(\hat{\bm\beta}_\alpha-\hat{\bm\beta}_{\alpha^*}^\alpha) \rightsquigarrow\mathbb{G}(\alpha)$ and $\Lambda_\alpha\rightsquigarrow q_\alpha=\mathbb{G}^\top(\alpha){\cal I}_\alpha\mathbb{G}(\alpha) $ for all functionals generated by $\alpha\in\{\alpha\cup\alpha^*: \alpha\in\Omega,\alpha\not\subseteq\alpha^*\}$.  
\end{lem}

We consider two scenarios. In the first, we assume that $p$ polynomially grows with $n$, meaning that there exists $c>0$ and $d>1$, such that $p/n^d\rightarrow c$ as $n\rightarrow\infty$. In the second, we assume that $p$ exponentially grows with $n$, meaning that there exist $c>0$ and $d\in(0,1/2)$ such that $\log{p}/n^d\rightarrow c$ as $n\rightarrow\infty$. 

We modify the method utilized by~\cite{fanlv2008} to show the sure screening property of the \textsf{ALRSIS} and \textsf{ASSIS} methods because we can treat ${\bm x}_j\circ{\bm x}_k$ as a main effect theoretically. The approach has been previously used for the theoretical properties of the all-pair SIS by~\cite{hallxue2014}. We do not need to consider any variables outside of the shrunk variable set in the penalization stage if consistency is reached. 

\begin{thm}
\label{thm:consistency of ALRSIS and ASSIS} (Sure screening property). 
Suppose that all assumptions of Lemma~\ref{lem:FCLT under high-dimensional setting} are satisfied, $|\alpha^*|=o(n/\log{n})$, and either (a) there exist $c>0$ and $d>1$ such that $\lim_{n\rightarrow\infty} p/n^d=c$ and $\mathop{\lim\!\inf}_{n\rightarrow\infty} n^v\delta_n>0$ for some $v<1/2$, or (b) there exist $c>0$ and $d\in(0,1/4)$ such that $\lim_{n\rightarrow\infty} \log p/n^d=c$ is satisfied and $\mathop{\lim\!\inf}_{n\rightarrow\infty} n^v\delta_n>0$ for some $v<1/4-d$. Then $\lim_{n\rightarrow\infty}P(\alpha_M\cup {\mathcal A}_{\alpha}^{LR}\supseteq \alpha_M^*)=1$ and $\lim_{n\rightarrow\infty}P(\alpha_M\cup {\mathcal A}_{\alpha}^{S}\supseteq \alpha_M^*)=1$ for any $\alpha\in\Omega$. 
\end{thm}

\begin{thm}
\label{thm:goodness of the fit statistic of the true model}
Suppose that Theorem~\ref{thm:consistency of ALRSIS and ASSIS} are satisfied. If $\lambda\ge [(4+\epsilon)/n]\log{p}$ for some $\epsilon>0$ or $\lambda=(1/n)\log{p}\log\!\log{n}$, then $\lim_{n\rightarrow\infty}{\rm Pr}\{|\hat\alpha_\lambda|\le |\alpha^*|+{n}\log{n}/(4\log{p})\}=1$, and $\lim_{n\rightarrow\infty}{\rm Pr}\{|\alpha|\le |\alpha^*|+ n/(4d)\}=1$ under (a) or $\lim_{n\rightarrow\infty}{\rm Pr}\{|\alpha|\le |\alpha^*|+ n^{1-4d}\log(n)/(4c)\}=1$ under (b).
\end{thm}

\begin{thm}
\label{thm:consistency when p polynomially increases with n}
(Consistency and oracle properties under the high-dimensional setting). Suppose that all assumptions of Theorem~\ref{thm:goodness of the fit statistic of the true model} are satisfied. If $\lambda=[(4+\epsilon)/n]\log{p}$ for some $\epsilon>0$ or $\lambda=(1/n)\log{p}\log\!\log{n}$, then $\hat\alpha_\lambda\stackrel{P}\rightarrow \alpha^*$ and the oracle properties are satisfied by~\eqref{eq:estimator of the L0 PML}.
\end{thm}

We have derived consistency and oracle properties of our method under the case when $\phi$ is present in~\eqref{eq:distribution of exponential family}. The corresponding conclusions hold under the case when $\phi$ is absent. In addition, we investigate our proofs and find that they only rely on the asymptotic properties of maximum likelihood estimation, implying that the corresponding conclusions can be extended to arbitrary high-dimensional statistical models mentioned in Section~\ref{subsec:extension}. We summarize those by the following.

\begin{cor}
\label{cor:extension to the case when phi is not present}
Suppose that $\phi$ is absent in the full HDGLM defined by~\eqref{eq:high-dimensional GLM with interaction effects}, implying that it is derived by assuming $\phi=1$ in~\eqref{eq:distribution of exponential family}. If the assumptions of Theorem~\ref{thm:consistency under the low-dimensional setting} hold, then the corresponding conclusions are satisfied under the low-dimensional setting. If the assumptions of Theorem~\ref{thm:consistency when p polynomially increases with n} hold, then the corresponding conclusions are satisfied under the high-dimensional setting.
\end{cor}

\begin{thm}
\label{thm:extension to general high-dimensional statistical models}
Suppose that the PDF/PMF of $y_i$ can be expressed as $f(y_i;\eta_i,\phi)$ with $\eta_i$ given by the right-hand side of~\eqref{eq:high-dimensional GLM with interaction effects}. If the assumptions of Theorem~\ref{thm:consistency under the low-dimensional setting} hold, then the corresponding conclusions are satisfied under the low-dimensional setting. If the assumptions of Theorem~\ref{thm:consistency when p polynomially increases with n} hold, then the corresponding conclusions are satisfied under the high-dimensional setting.
\end{thm}

We examine the local optimizers given by Algorithm~\ref{alg:two stage algorithm for interaction effects}. We investigate the theoretical properties of the addition and the removal steps, respectively. In the addition step, if there exists $j\in\alpha_M^*$ but $j\not\in\alpha_M$ or $(j,k)\in\alpha_I^*$ but $(j,k)\not\in\alpha_I$, then it is unlikely to expand $\alpha$ by adding a main or an interaction effect not contained $\alpha^*$. In the removal step, if there exists $j\in\alpha$ but $j\not\in\alpha_M^*$ or $(j,k)\in\alpha_I$ but $(j,k)\not\in\alpha_M^*$, then it is unlikely to shrink $\alpha$ by removing a main or an interaction effect contained by $\alpha^*$. Therefore, we expect that $|\alpha\cap \alpha^*|$ grows in the iterations. We only provide our theorems for the high-dimensional setting in the following.

\begin{thm}
\label{thm:local optimizers}
If all assumptions of Theorem~\ref{thm:consistency when p polynomially increases with n} are satisfied, then any local solution of $\hat\alpha_\lambda$ provided by Algorithm~\ref{alg:two stage algorithm for interaction effects} converges to $\alpha^*$ in probability and the corresponding local solutions of $\hat{\bm\beta}_\lambda$ and $\hat\phi_\lambda$ enjoy oracle properties.
\end{thm}

\begin{cor}
\label{cor:finite step iterations}
If all assumptions of Theorem~\ref{thm:consistency when p polynomially increases with n} are satisfied and the upper bound of the iterations is set to be $|\alpha^*|^2$, then the final solution of  $\hat\alpha_\lambda$ provided by Algorithm~\ref{alg:two stage algorithm for interaction effects} converges to $\alpha^*$ in probability and the corresponding local solutions of $\hat{\bm\beta}_\lambda$ and $\hat\phi_\lambda$ enjoy oracle properties.
\end{cor}

\section{Simulation}
\label{sec:simulation}

We compared our proposed method with our competitors via Monte Carlo simulations with $1000$ replications. We assumed that variable selection was carried out under the full model given by~\eqref{eq:high-dimensional GLM with interaction effects} with~\eqref{eq:distribution of exponential family} specified for the HDLM for a normal response given by~\eqref{eq:full model with interaction effects} and the HDLOGIT for a binomial response given by~\eqref{eq:ultrahigh-dimensional logistic model}, respectively.  In both the models, we applied the \textsf{ASSIS} method with $\gamma=1/\log{n}$ in the screening stage. After the shrunk variable set was derived, we chose $\lambda=(1/n)\log{p}\log\!\log{n}$ in the penalization stage to select the main effects and interactions. 

In our simulation for the HDLM given by~\eqref{eq:full model with interaction effects}, we chose $n=200$ and $p=2000$. We generated rows of variable matrix ${\bf X}_M=[{\bm x}_1,\dots,{\bm x}_p]$ independently. We incorporated dependencies between the columns by normal distributions with ${\rm cov}(x_{ij},x_{ik})=\rho^{|\tau(j)-\tau(k)|}$ and $\rho=0.0,0.5,0.8$ for $i=1,\dots,n$ and $j,k=1,\dots,p$, respectively, where $\tau$ was a random permutation on $\{1,\dots,p\}$. After ${\bf X}_M$ was derived, we generated the $200$-dimensional response ${\bm y}$ from the true model as
\begin{equation}
\label{eq:true linear model simulation}
\eqalign{
{\bm y}=&\beta_1^*{\bm x}_1+\beta_2^*{\bm x}_2+\beta_3^*{\bm x}_3+\beta_4^*{\bm x}_4+\beta_5^*{\bm x}_5+\beta_6^*{\bm x}_6\cr
&+\beta_{14}^*{\bm x}_1\circ{\bm x}_4+\beta_{15}^*{\bm x}_1\circ{\bm x}_5+\beta_{56}^*{\bm x}_5\circ{\bm x}_5+{\bm\epsilon}^*, {\bm\epsilon}^*\sim{\mathcal N}(0,{\bf I}).
}
\end{equation}
We fixed $\beta_{14}^*=\beta_{15}^*=\beta_{16}^*=3$, leading to $\alpha_{I}^*={\mathcal I}^*=\{(1,4),(1,5),(5,6)\}$. We considered three cases of the main effects. In Case (a), we chose $\beta_1^*=\beta_2^*=\beta_3^*=\beta_4^*=3$ and $\beta_5^*=\beta_6^*=0$, leading to ${\mathcal M}^*=\{1,2,3,4\}$ and $\alpha_M^*=\{1,2,3,4,5,6\}$. In Case (b), we chose $\beta_1^*=\beta_2^*=\beta_3^*=\beta_4^*=\beta_5^*=\beta_6^*=3$, leading to ${\mathcal M}^*=\alpha_M^*=\{1,2,3,4,5,6\}$. In Case (c), we chose $\beta_1^*=\beta_2^*=\beta_3^*=\beta_4^*=\beta_5^*=\beta_6^*=0$,  leading to ${\mathcal M}^*=\emptyset$ and $\alpha_M^*=\{1,4,5,6\}$.

\begin{table}
\tiny
\scriptsize
\caption{\label{tab:simulation for regression models} Simulation with $1000$ replications for percentages ($\%$) of true positives and numbers ($\#$) of false positives for the main (M) effects and interactions (I), respectively, provided by our proposed \textsf{SHL0} method under the true HDLM defined by~\eqref{eq:true linear model simulation} or the true HDLOGIT defined by~\eqref{eq:true logistic model simulation}.}
\begin{center}
\begin{tabular}{ccccccccccc}\hline
                         &&  \multicolumn{3}{c}{$\rho=0.0$} &  \multicolumn{3}{c}{$\rho=0.5$} &  \multicolumn{3}{c}{$\rho=0.8$}  \\\cline{3-11}
                         && (a) & (b) & (c)  & (a) & (b) & (c)  & (a) & (b) & (c) \\\hline
                         && \multicolumn{9}{c}{\% of True Positives for~\eqref{eq:true linear model simulation}} \\
\textsf{SHL0} & M& $99.8$ & $99.1$ & $100.0$ & $99.8$ & $98.3$ & $100.0$ & $99.1$ & $96.8$ & $100.0$  \\
     & I & $99.8$ & $99.1$ & $100.0$ & $99.8$ & $97.8$ & $100.0$ & $98.8$ & $95.8$ & $100.0$ \\
\textsf{GRESH} & M & $99.8$ & $98.7$ & $100.0$ & $99.8$ &  $98.1$ & $100.0$ & $98.5$ & $94.8$ & $98.0$   \\
   & I &  $99.7$ & $98.7$ & $100.0$ & $99.7$ & $97.4$ & $100.0$ & $94.8$ & $94.0$ & $98.0$ \\
\textsf{SHIM} & M & $89.9$ &  $99.0$ & $44.7$ & $88.6$ & $98.3$ &  $32.6$ & $74.7$ & $94.9$ & $39.5$    \\
  & I &  $80.6$ & $99.1$ & $44.4$ & $78.7$ & $97.4$ & $48.6$ & $55.5$ & $93.5$ & $38.8$\\
\textsf{hierNet} & M & $5.3$ & $6.3$ & $65.6$ &  $4.9$ & $5.8$ & $65.4$ & $5.3$ & $5.3$ & $66.6$   \\
  & I &  $0.1$ & $0.2$ & $3.7$ & $0.0$ & $0.0$ & $3.5$ & $0.2$ & $0.3$ & $7.6$ \\
\textsf{RAMP} &  M & $68.3$ & $100.0$ & $0.4$ & $68.2$ & $100.0$ & $0.4$ & $66.6$ & $99.1$ & $0.2$\\
  & I &  $36.3$ & $99.9$ & $0.2$ & $35.5$ & $99.9$ &  $0.2$ & $33.4$ & $96.8$ & $0.1$ \\
\textsf{hierScale} & M & $60.8$ & $91.9$ & $52.0$ & $64.7$ & $91.2$ & NA & $58.5$ & $93.1$ & NA \\
 & I &   $11.9$ & $72.9$ & $46.8$ & $17.9$ & $71.9$ & NA & $10.8$ & $92.1$ & NA\\
\textsf{HiQR} & M & $66.5$ & $96.7$ & $0.0$ & $66.6$ & $95.8$ & $0.0$ & $66.3$ & $93.6$ &  $0.0$ \\
 & I &  $99.6$ & $96.8$ & $100.0$ & $98.7$ & $95.8$ & $100.0$ & $99.2$ & $93.7$ & $100.0$ \\
\textsf{sprinter} & M & $0.3$ & $0.0$ & $0.0$ & $0.2$ & $0.1$ & $0.0$ & $0.3$ & $0.4$ & $0.0$     \\
   & I &  $0.0$ & $0.0$ & $0.0$ & $0.0$ & $0.0$ & $0.0$ & $0.0$ & $0.0$ & $0.0$ \\
                         & \multicolumn{9}{c}{\# of False Positives for~\eqref{eq:true linear model simulation}} \\ 
\textsf{SHL0} & M& $0.003$ & $0.000$ & $0.053$ & $0.004$ & $0.011$ & $0.038$& $0.036$ &  $0.116$ & $0.054$  \\
     & I & $0.001$ & $0.002$ & $0.022$ & $0.008$ & $0.011$ & $0.019$ & $0.023$ & $0.078$ & $0.009$ \\
\textsf{GRESH} & M & $1.995$ & $1.844$ & $2.246$ & $2.050$ &  $2.008$ &$2.398$ & $3.747$ &$3.724$ & $3.462$  \\
   & I &  $0.367$ & $0.315$ & $0.373$ & $0.537$ &  $0.395$ & $0.653$ & $3.702$ & $3.318$ & $3.169$  \\
\textsf{SHIM} & M & $1.132$ &$0.142$ & $2.351$ &  $1.411$ & $0.364$ & $2.734$ & $1.554$ & $2.168$ & $2.571$  \\
  & I &  $4.882$ & $6.998$ &  $0.671$ & $4.711$ & $6.731$ & $0.625$ & $2.968$ & $4.262$ & $0.958$ \\
\textsf{hierNet} & M & $0.886$ & $0.927$ & $3.994$ &$0.916$ & $0.941$ & $4.012$ & $0.961$ & $1.017$ & $4.372$  \\
  & I & $1.001$ & $1.007$ &  $3.182$ & $1.008$ & $1.009$ & $3.214$ & $1.004$ & $1.013$ & $3.673$ \\
\textsf{RAMP} &  M & $0.575$ & $0.083$ & $1.001$ & $0.566$ & $0.099$ & $1.020$ & $0.499$ & $0.346$ & $0.995$ \\
  & I &  $0.102$ & $0.004$ & $0.999$ & $0.101$ & $0.008$ & $0.996$ & $0.160$ & $0.161$ & $0.998$ \\
\textsf{hierScale} & M & $0.428$ & $3.709$ & $8.670$ & $1.004$ & $3.724$ & NA & $1.035$ & $6.729$ & NA  \\
 & I &   $0.002$ & $0.005$ & $0.160$ & $0.011$ & $0.018$ & NA & $0.107$ & $0.361$ & NA \\
\textsf{HiQR} & M & $0.023$ & $0.067$ & $1.000$ & $0.069$ &  $0.166$ & $1.000$ & $0.582$ & $1.181$ & $1.000$ \\
 & I & $2.276$ & $3.100$ &  $0.697$ & $2.224$ &  $3.131$ & $0.698$ & $2.385$ & $3.118$ & $1.180$ \\
\textsf{sprinter} & M & $4.516$ & $5.911$ & $1.109$ & $4.519$ &$5.937$ & $1.083$ &$5.090$ &  $6.899$ & $1.120$   \\
   & I &  $4.613$ & $5.396$ & $3.796$ & $4.590$ & $5.286$ &  $3.801$ & $4.788$ & $5.316$ & $4.061$ \\\hline
                         & \multicolumn{9}{c}{\% of True Positives for~\eqref{eq:true logistic model simulation}} \\ 
\textsf{SHL0} & M & $99.8$ & $100.0$ & $99.7$   & $99.8$ & $100.0$ & $99.7$ & $99.7$ & $100.0$ & $99.5$ \\
  & I & $100.0$ & $100.0$ &  $100.0$ & $100.0$ & $100.0$ &$100.0$  & $100.0$ & $100.0$ &$100.0$ \\
\textsf{SODA} &   M &  $50.0$ & $100.0$ & NA  & $50.0$ & $99.6$ & NA & $49.3$ & $93.8$  & NA  \\
 & I & $99.2$ &  $99.9$ & NA & $98.3$ & $99.4$ & NA  & $99.0$ & $88.4$ & NA \\
                         & \multicolumn{9}{c}{\# of False Positives for~\eqref{eq:true logistic model simulation}} \\ 
\textsf{SHL0} &  M &  $0.007$ & $0.009$ & $0.009$ & $0.011$ & $0.006$ & $0.011$ & $0.006$ & $0.009$ & $0.005$ \\
   & I & $0.000$ & $0.000$ & $0.000$  & $0.002$ & $0.000$ &  $0.000$ & $0.000$ & $0.000$ & $0.000$ \\
\textsf{SODA} & M & $0.000$ &  $0.000$ &NA  & $0.000$ &  $0.000$ & NA  & $0.004$ & $0.033$  & NA \\
& I &  $0.000$ & $0.000$ &  NA & $0.015$ & $0.002$ &  NA &  $0.183$ & $0.159$ & NA \\\hline
\end{tabular}
\end{center}
\end{table}

We applied the proposed \textsf{SHL0} based on Algorithm~\ref{alg:two stage algorithm for interaction effects}. The results were compared with the previous \textsf{GRESH}~\cite{she2018}, \textsf{SHIM}~\cite{choi2010}, \textsf{hierNet}~\cite{bien2013},  \textsf{HiQR}~\cite{wang2024}, \textsf{sprinter}~\cite{yu2023},  \textsf{RAMP}~\cite{hao2018}, and \textsf{hierScale}~\cite{hazimeh2020} methods based on data generated from~\eqref{eq:true linear model simulation}. The previous \textsf{GRESH}, the \textsf{SHIM}, and the \textsf{hierNet} method could not applied to the full model with $p$ main effects and $p(p-1)/2$ interactions. We combined those with the \textsf{ASSIS}. They were applied to the same shrunk variable set provided by the screening stage of our method. We did not use the \textsf{ASSIS} to screen variables in the \textsf{HiQR}, the \textsf{sprinter}, the \textsf{RAMP}, and the \textsf{hierScale} methods because they have their built-in SIS. The built-in SIS used in the \textsf{HiQR} and \textsf{sprinter} is the all-pairs SIS proposed by~\cite{hallxue2014} but a concern is that the all-pair SIS may violate the SH restriction. The built-in SIS in the \textsf{RAMP} selects the main effects first. It then considers their associated interactions. A concern is that this approach may not capture important interactions if the associated main effects are weak or absent. The built-in SIS adopted by the \textsf{hierScale} is developed based on the proximal gradient descent (PGD). The criterion is based on a summation but not a maximization, leading to a similar concern in the \textsf{RAMP}. 

We calculated the numbers of true and false positives for the main and interaction effects for the eight methods, respectively. The true and false positive sets for the main effects were $TP_M=\{j:j\in\hat\alpha_M, j\in\alpha_M^*\}$ and $FP_M=\{j:j\in\hat\alpha_M,j\not\in\alpha_M^*\}$, respectively. The true and false positive sets for the interaction effects were computed by $TP_I=\{(j,k): (j,k)\in\hat\alpha_I,(j,k)\in\alpha_I^*\}$ and $FP_I=\{(j,k):(j,k)\in\hat\alpha_I,(j,k)\not\in\alpha_I^*\}$, respectively. After they were derived, we computed the proportions of true positives of the main effects by $|TP_M|/6$ in Cases (a) and (b) and $|TP_M|/4$ in Case (c), respectively, and the proportions of true positives for the interaction effects by $|TP_I|/3$ in all the three cases (Table~\ref{tab:simulation for regression models}, the first half for~\eqref{eq:true linear model simulation}). We also computed the number of false positives of the main and the interactions by $|FP_M|$ and $|FP_I|$, respectively (Table~\ref{tab:simulation for regression models}, the second half for~\eqref{eq:true linear model simulation}). 

The simulation showed that the proposed \textsf{SHL0} had the greatest true positive rates and the least false positive values, indicating that it was the best. The \textsf{GRESH} was the second best because, with a high probability, it could capture all the important main effects and interactions. The \textsf{SHIM} could capture important interaction effects if their associated main effects were important but not otherwise. The \textsf{hierNet} method missed many important main effects and interactions. It was unlikely to use the \textsf{HiQR} and \textsf{sprinter} methods to capture the associated main effects even if they were related to important interactions. It was hard to use the \textsf{RAMP} and \textsf{hierScale} methods to capture important interactions if their related main effects were weak or absent because they missed ${\bm x}_5\circ{\bm x}_6$ in Case (c) of the simulation. The simulation study showed the proposed \textsf{SHL0} and the previous \textsf{GRESH} were appropriate for selecting important interactions but not the other six methods. The performance of the \textsf{SHL0} was better than the previous \textsf{GRESH}.

For ultrahigh-dimensional logistic models, we chose $n=500$ and $p=100$ with the same way to generate ${\bf X}_M=[{\bm x}_1,\dots,{\bm x}_p]$. After ${\bf X}_M$ was derived, we generated the $500$-dimensional response from the true model as ${\bm y}\sim {\mathcal Bin}(10,{\bm\pi}^*)$ with 
\begin{equation}
\label{eq:true logistic model simulation}
\eqalign{
\log{{\bm\pi}^*\over{\bm 1}-{\bm\pi}^*}=&\beta_1^*{\bm x}_1+\beta_2^*{\bm x}_2+\beta_3^*{\bm x}_3+\beta_4^*{\bm x}_4+{\bm\beta}_{12}^*{\bm x}_1\circ{\bm x}_2+{\bm\beta}_{13}^*{\bm x}_1\circ{\bm x}_3+{\bm\beta}_{34}^*{\bm x}_3\circ{\bm x}_4.
}
\end{equation}
We fixed $\beta_{12}^*=\beta_{13}^*=\beta_{34}^*=3.0$. We considered three cases of the main effects. In Case (a), we chose $\beta_1^*=\beta_2^*=3$ and $\beta_3^*=\beta_4^*=0$, leading to ${\mathcal M}^*=\{1,2\}$, and $\alpha_M^*=\{1,2,3,4\}$. In Case (b), we chose $\beta_1^*=\beta_2^*=\beta_3^*=\beta_4^*=3$, leading to $\alpha_M^*={\mathcal M}^*=\{1,2,3,4\}$. In Case (c), we chose $\beta_1^*=\beta_2^*=\beta_3^*=\beta_4^*=0$,  leading to ${\mathcal M}^*=\emptyset$, and $\alpha_M^*=\{1,2,3,4\}$. 

We compared our proposed \textsf{SHL0} with the previous \textsf{SODA}~\cite{li2019} methods based on data generated from~\eqref{eq:true logistic model simulation}. It showed that the proposed \textsf{SHL0} was always better than the previous \textsf{SODA}. We encountered more than $50\%$ of computational errors in Case (c) when the previous \textsf{SODA} was applied, implying that it could not be used if the interactions were important but the associated main effect were not. 

\begin{table}
\caption{\label{tab:violations of strong hierarchical restrictions} Simulations with $1000$ replications for percentages of violations of the SH restriction in our proposed \textsf{SHL0} and the previous \textsf{GRESH}, \textsf{SHIM}, \textsf{hierNet}, \textsf{RAMP}, \textsf{hierScale}, \textsf{HiQR}, \textsf{sprinter},  and \textsf{SODA} methods.}
\begin{center}
\begin{tabular}{ccccccccccc}\hline
    & \multicolumn{3}{c}{$\rho=0.0$} & \multicolumn{3}{c}{$\rho=0.5$} & \multicolumn{3}{c}{$\rho=0.8$} \\\cline{2-10}
  & (a) & (b) & (c) & (a) & (b) & (c) & (a) & (b) & (c)  \\\hline
           & \multicolumn{9}{c}{\% of Violations for~\eqref{eq:true linear model simulation}} \\
\textsf{SHL0} & $0.0$ & $0.0$ & $0.0$ & $0.0$ & $0.0$ & $0.0$ & $0.0$ & $0.0$ & $0.0$   \\
\textsf{GRESH} &   $0.0$ & $0.0$ & $0.0$ & $0.0$ & $0.0$ & $0.0$ & $0.0$ & $0.0$ & $0.0$\\
\textsf{SHIM} &   $0.0$ & $0.0$ & $0.0$ & $0.0$ & $0.0$ & $0.0$ & $0.0$ & $0.0$ & $0.0$\\
\textsf{hierNet} &   $0.0$ & $0.0$ & $0.0$ & $0.0$ & $0.0$ & $0.0$ & $0.0$ & $0.0$ & $0.0$\\
\textsf{RAMP} &   $0.0$ & $0.0$ & $0.0$ & $0.0$ & $0.0$ & $0.0$ & $0.0$ & $0.0$ & $0.0$\\
\textsf{hierScale} & $0.0$ & $0.0$ & $0.0$ & $0.0$ & $0.0$ & $0.0$ & $0.0$ & $0.0$ & $0.0$ \\
\textsf{HiQR} & $99.7$ & $89.4$ & $100.0$ & $99.9$  & $88.7$& $100.0$ & $99.9$  & $86.2$ & $100.0$  \\
\textsf{sprinter} &$98.5$ & $87.3$ & $99.9$ & $98.8$ & $85.4$ & $100.0$ & $99.0$ & $80.0$ &  $100.0$  \\
            & \multicolumn{9}{c}{\% of Violations for~\eqref{eq:true logistic model simulation}} \\
\textsf{SHL0} & $0.0$ & $0.0$ & $0.0$ & $0.0$ & $0.0$ & $0.0$ & $0.0$ & $0.0$ & $0.0$   \\
\textsf{SODA} & $100.0$ & $0.0$ & NA & $100.0$ & $0.0$ & NA & $99.2$ & $0.4$ & NA     \\\hline
\end{tabular}
\end{center}
\end{table}

We next evaluated violations of the SH restriction. It showed that the SH restriction was likely to be violated in the previous \textsf{HiQR}, the \textsf{sprinter}, and the \textsf{SODA} methods. The three methods should be treated as inappropriate in selecting the main effects and the interactions because the SH restriction is required based on the statistical literature. 

In summary, our simulation study shows that the proposed \textsf{SHL0} method outperforms its competitors in both the HDLM and the HDLOGIT models, indicating that it is the best. The proposed \textsf{ASSIS} method can be combined with other variable selection methods for the interaction screening. Examples include the previous \textsf{GRESH}, the \textsf{SHIM}, and the \textsf{hierNet} methods. It is inappropriate to use the previous all-pair SIS to screen interactions because it can induce a violation of the SH restriction. A few previous methods may violate the SH restriction. Examples include the previous \textsf{HiQR}, the \textsf{sprinter}, and the \textsf{SODA}. 

\section{Application}
\label{sec:application}

We applied our method to the Prostate Cancer data set provided by~\cite{singh2002}. It is a gene expression microarray data set of $12,\!600$ genes of $77$ prostate cancer patients and $59$ normal specimens. Prostate cancer is one of the four most common cancers in the United States with about $288,\!300$ new cases and $34,\!700$ deaths in estimates of $2023$. It has been pointed out that cancer screening has led to a significant reduction in the death rate of prostate cancer. The challenge is to identify those patients at risk for relapse with a better understanding of the molecular abnormalities of tumors at risk for relapse. The goal of the microarray data is to address the challenge. We compared our proposed method with the previous \textsf{SODA} method. We also studied the possibility of the remaining methods in Table~\ref{tab:simulation for regression models}. We did not consider those methods because they are developed for the HDLMs for normal responses but not an ultrahigh-dimensional logistic model for Bernoulli responses. 

The response ${\bm y}$ is $136$-dimensional Bernoulli random vector with $y_i=1$ represent a prostate cancer patient or $y_i=0$ otherwise, $i=1,\dots,136$. The size of the variable matrix is $136\times 12,\!600$, implying that we have $n=136$ and $p=12,\!600$ in our data. We use the logistic linear model for the Bernoulli response to analyze the data, leading to the full model as
\begin{equation}
\label{eq:full model logistic prostate data}
\log{{\bm \pi}\over{\bm 1}-{\bm\pi}}=\beta_0+\sum_{j=1}^{12600} {\bm x}_j\beta_j+\sum_{j=1}^{12599}\sum_{k=j+1}^{12600}{\bm x}_j\circ{\bm x}_k\beta_{jk},
\end{equation}
where ${\bm\pi}={\rm E}({\bm y})$. The design matrix of the full model given by~\eqref{eq:full model logistic prostate data} has $79,\!386,\!301$ columns and $136$ rows, including an intercept, $12,\!600$ main effects, and $79,\!373,\!700$ interactions. As too many columns are contained, it is important to use a variable selection procedure to remove unimportant main effects and interactions from the right-hand side of~\eqref{eq:full model logistic prostate data} to interpret the response better.

 We applied our method to select the main effects and interactions under the SH restriction. We chose the EBIC given by $\kappa=\log{p}\log\!\log{n}=15.03$ in the GIC to select the tuning parameter. We obtained the main effects ${\bm x}_{6185}$ and ${\bm x}_{6366}$, and the interaction ${\bm x}_{6185}\circ{\bm x}_{6366}$. The residual deviance of the selected model was $G^2=48.90$. Compared it with the residual deviance of the null model $G_{null}^2=186.15$, our method explained about $73.7\%$ of the total variations of the data (i.e., $(186.15-48.90)/186.15=73.7\%$). In addition, we evaluated three related models. The first was obtained by excluding the interaction from the selected model, leading to the model with main effects ${\bm x}_{6185}$ and ${\bm x}_{6366}$. The residual deviance of the first related model was $G_1^2=75.56$. Because $G_1^2-G^2>15.03$, our method preferred the selected but not the first related model. The second was obtained by removing ${\bm x}_{6185}$ from the selected model, leading to the model with the main effect ${\bm x}_{6366}$ and the interaction ${\bm x}_{6185}\circ{\bm x}_{6366}$. The residual deviance of the second related model was $G_2^2=123.48$. Note that it violated the SH restriction and $G_2^2-G^2=74.58>15.03$. We preferred the selected model but not the second related model even if the SH restriction was ignored. The third was obtained by removing ${\bm x}_{6366}$ from the selected model, leading to the model with the main effect ${\bm x}_{6185}$ and the interaction ${\bm x}_{6185}\circ{\bm x}_{6366}$. The residual deviance of the third related model was $G_3^2=49.11$. Because it violated the SH restriction and $G_3^2-G^2=0.21<15.03$, our method would prefer the third related model if the SH restriction were ignored. We reported the selected but not the third related model because our method considered the SH restriction. 

To compare, we also applied the \textsf{SODA} method. The \textsf{SODA} selected the main effect ${\bm x}_{10672}$ and the interaction ${\bm x}_{10234}\circ{\bm x}_{10672}$. The residual deviance of the selected model was $G^2=179.65$. The \textsf{SODA} violated the SH restriction because it did not include the main effect ${\bm x}_{10234}$. We then studied a related model derived by adding ${\bm x}_{10234}$ to the selected model, such that it satisfied the SH restriction. The residual deviance of the related model was $G_1^2=177.91$. The selected model reported by the \textsf{SODA} explained about $3.5\%$ of the total variations. The related model explained about $4.4\%$.  Based on the comparison, we concluded that the performance of our method was better than the \textsf{SODA} in the application. 

\section{Conclusion}
\label{sec:conclusion}

The proposed \textsf{SHL0} is a convenient method for selecting the main effects and interactions from an arbitrary ultrahigh-dimensional statistical model. The computation is easy because it only needs a maximum likelihood algorithm. It is unnecessary to devise an algorithm for a global solution. The nondeterministic approach of the local combinatorial optimization may induce a global optimizer of the objective function. It is a nice property of the $L_0$ penalty for selecting interactions from an ultrahigh-dimensional statistical model. 


\appendix
\section{Proofs}
\label{sec:proofs}

\noindent
{\bf Proof of Threorem~\ref{thm:formulation for the aggregated score statistic}.} Denote ${\bf X}_{jk}=({\bf X}_\alpha,{\bm x}_{jk})$, ${\bm x}_{jk}={\bm x}_{j}\circ{\bm x}_k$, ${\bm\beta}_{jk}=({\bm\beta}_\alpha^\top,\beta_{jk})^\top$, ${\bm z}_{jk}={\bf X}_{jk}{\bm\beta}_{jk}+({\bm y}-{\bm\mu}_{\alpha\cup\{(j,k)\}})\circ {\partial{\bm\eta}_{\alpha\cup\{(j,k)}}/\partial{\bm\mu}_{\alpha\cup\{(j,k)\}}$,  and ${\bf W}_{jk}={\rm diag}\{(\partial{\bm\mu}_{\alpha\cup\{(j,k)\}}/\partial{\bm\eta}_{\alpha\cup\{(j,k)\}} )^2/b''({\bm\theta}_{\alpha\cup\{(j,k)\}})\}$. By ${\rm var}(y_i)=\phi b''(\theta_i)=\phi{\partial\mu_i}/\partial\theta_i=\phi({\partial\mu_i}/\partial\eta_i)(\partial\eta_i/\partial\theta_i)$, we obtain $\partial{\bm\theta}_{\alpha\cup\{(j,k)\}}/\partial{\bm\eta}_{\alpha\cup\{(j,k)\}}={\bf W}_{jk}{\partial{\bm\eta} _{\alpha\cup\{(j,k)\}}/\partial{\bm\mu}_{\alpha\cup\{(j,k)\}}}$. The gradient vector of the loglikelihood function of~\eqref{eq:expanded model interaction j and k} with respect to ${\bm\beta}_{jk}$ is
$$\eqalign{
\dot\ell({\bm\beta}_{jk},\phi)=&{\partial\ell({\bm\beta}_{jk},\phi)\over\partial{\bm\beta}_{jk}}={1\over\phi}{\bf X}_{jk}^\top({\bm y}-{\bm\mu}_{\alpha\cup\{(j,k)\}})\circ{\partial{\bm\theta}_{\alpha\cup\{(j,k)\}}\over\partial{\bm\eta}_{\alpha\cup\{(j,k)\}}}\cr
=&{1\over\phi}{\bf X}_{jk}^\top {\bf W}_{jk} ({\bm z}_{jk}-{\bf X}_{jk}{\bm\beta}_{jk}) .\cr
}$$
The Hessian matrix with respect to ${\bm\beta}_{jk}$ is
$$\eqalign{
\ddot\ell({\bm\beta}_{jk},\phi)
=&-{1\over\phi}{\bf X}_{jk}^\top{\bf W}_{jk}{\bf X}_{jk}+ {1\over\phi}{\bf X}_{jk}^\top{\rm diag}\left\{({\bm y}-{\bm\mu}_{\alpha\cup\{(j,k)\}})\circ{\partial^2{\bm\theta}_{\alpha\cup\{(j,k)\}}\over\partial{\bm\eta}_{\alpha\cup\{(j,k)\}}^2}\right\}{\bf X}_{jk}.\cr
}$$
Because the excepted value of the second term is ${\bm 0}$, we obtain the Fisher Information matrix as $I({\bm\beta}_{jk})=\phi^{-1}{\bf X}_{jk}^\top{\bf W}_{jk}{\bf X}_{jk}$. Substituting $\breve{\bm\beta}_{\alpha\cup\{(j,k)\}}$ for ${\bm\beta}_{jk}$, we obtain ${\bf W}_{jk}={\bf W}$, ${\bf X}_{jk}\breve{\bm\beta}_{\alpha\cup\{(j,k)\}}={\bf X}_\alpha\hat{\bm\beta}_\alpha$, ${\bm z}_{jk}={\bm z}_\alpha$, 
$$\eqalign{
\phi\dot\ell(\breve{\bm\beta}_{\alpha\cup\{(j,k)\}})=&\left(\begin{array}{c} {\bf X}_\alpha^\top \cr {\bm x}_{jk}^\top \cr  \end{array}  \right){\bf W}_\alpha({\bm z}_\alpha-{\bf X}_\alpha\hat{\bm\beta}_\alpha)=\left( \begin{array}{c}  {\bf X}_\alpha^\top{\bf W}_\alpha({\bm z}_\alpha-{\bf X}_\alpha\hat{\bm\beta}_\alpha) \cr {\bm x}_{jk}^\top {\bf W}_\alpha({\bm z}_\alpha-{\bf X}_\alpha\hat{\bm\beta}_\alpha) \cr   \end{array}   \right)
}$$
and 
$$\eqalign{
\phi I^{-1}(\breve{\bm\beta}_{\alpha\cup\{(j,k)\}})=&\left[\left(\begin{array}{c} {\bf X}_\alpha^\top \cr {\bm x}_{jk}^\top \cr  \end{array}  \right) {\bf W}_\alpha   \left(\begin{array}{cc} {\bf X}_\alpha &  {\bm x}_{jk}  \end{array}  \right) \right]^{-1}\cr
=&\left(\begin{array}{cc} {\bf X}_\alpha^\top {\bf W}_\alpha{\bf X}_\alpha  &  {\bf X}_\alpha^\top{\bf W}_\alpha{\bm x}_{jk} \cr  {\bm x}_{jk}^\top{\bf W}_\alpha{\bf X}_\alpha &  {\bm x}_{jk}^\top{\bf W}_\alpha{\bm x}_{jk} \cr     \end{array}    \right)^{-1} \cr
=& {\bf A}^\top \left(   \begin{array}{cc} ({\bf X}_\alpha^\top {\bf W}_\alpha{\bf X}_\alpha)^{-1} &  {\bf 0} \cr {\bf 0} & \|{\bf I}-{\bf P}{\bf W}_\alpha^{1/2}{\bm x}_{jk}\|^{-2}\cr     \end{array} \right) {\bf A}, \cr
}$$
where
$${\bf A}=\left(\begin{array}{cc} {\bf I} & {\bf 0} \cr -{\bm x}_{jk}{\bf W}_\alpha {\bf X}_\alpha({\bf X}_\alpha^\top {\bf W}_\alpha{\bf X}_\alpha)^{-1}  & {\bf I}\end{array} \right).$$
By $\dot\ell({\bm\beta}_\alpha,\phi)=0$, we obtain $\hat{\bm\beta}_\alpha= ({\bf X}_\alpha^\top {\bf W}_\alpha{\bf X}_\alpha)^{-1}{\bf X}_\alpha^\top{\bf W}_\alpha{\bm z}_\alpha$, leading to $({\bf X}_\alpha^\top {\bf W}_\alpha{\bf X}_\alpha)^{-1}{\bf X}_\alpha^\top{\bf W}_\alpha({\bm z}_\alpha-{\bf X}_\alpha\hat{\bm\beta}_\alpha)={\bf 0}$ and 
$$
\phi{\bf A}\dot\ell(\breve{\bm\beta}_{\alpha\cup\{(j,k)\}})=\left(\begin{array}{c} {\bf X}_\alpha^\top{\bf W}_\alpha({\bm z}_\alpha-{\bf X}_\alpha\hat{\bm\beta}_\alpha) \cr {\bm x}_{jk}^\top {\bf W}_\alpha({\bm z}_\alpha-{\bf X}_\alpha\hat{\bm\beta}_\alpha) \cr     \end{array} \right).
$$
Because ${\bf P}_\alpha$ is an orthogonal matrix, we obtain 
$$\eqalign{
{\bm x}_{jk}^\top {\bf W}_\alpha({\bm z}_\alpha-{\bf X}_\alpha\hat{\bm\beta}_\alpha)=&{\bm x}_{jk}^\top{\bf W}_\alpha[{\bf I}- ({\bf X}_\alpha^\top {\bf W}_\alpha{\bf X}_\alpha)^{-1}{\bf X}_\alpha^\top{\bf W}_\alpha ]{\bm z}_\alpha\cr
=&\langle ({\bf I}-{\bf P}_\alpha){\bf W}_\alpha^{1/2}{\bm x}_{jk},  ({\bf I}-{\bf P}_\alpha){\bf W}_\alpha^{1/2}({\bm z}_\alpha-{\bf X}_\alpha{\bm\beta}_\alpha) \rangle,\cr
}$$
 leading to 
$$\eqalign{
S_{(j,k)|\alpha}
=&{\hat\phi_\alpha\langle  {\bm r}_\alpha,{\bm s}_{(j,k)|\alpha} \rangle \over \|{\bm s}_{(j,k)|\alpha}\|}.
}$$
We conclude. \qed

\noindent
{\bf Proof of Theorem~\ref{thm:strong hierarchy satisfied}.} If there exists $k\not=j$ such that $\beta_{jk}$ is contained in the resulting model determined by~\eqref{eq:estimator of the L0 PML}, then $I(\beta_{jk}\not=0)=1$ implying that ${\bm\xi}_j\not={\bm 0}$ leading to $I({\bm\xi}_j\not={\bm 0})$. As $\beta_{jk}$ is included, the value of $\|{\bm\xi}_j\|_{0,mod}$ does not change as $\beta_j$ varies. Therefore, the estimate of $\beta_j$ is completely determined by the first term on the right-hand side of~\eqref{eq:objective function L0 interaction effect}, implying that with probability one it is nonzero and the strong hierarchy is satisfied by $\hat\alpha_\lambda$. The penalty terms do not vary if estimation of ${\bm\beta}$ and $\phi$ is evaluated under $\hat\alpha_\lambda$, implying that $\hat{\bm\beta}_\lambda$ and $\hat\phi_\lambda$ are identical to the corresponding MLEs. We conclude. \qed

\noindent
{\bf Proof of Theorem~\ref{thm:consistency of L0 penalty}.}  The conclusion is directly implied by Theorems~\ref{thm:consistency under the low-dimensional setting} and~\ref{thm:consistency when p polynomially increases with n} of Section~\ref{sec:theoretical properties}. \qed

\noindent
{\bf Proof Lemma~\ref{lem:monotonicity of GIC}.} Large $\lambda$ induces less number of nonzero estimates of regression coefficients given by~\eqref{eq:estimator of the L0 PML}. If $|\hat\alpha_\lambda|$ does not change as $\lambda$ varies, then the penalty component contained by the right-hand side of~\eqref{eq:objective function L0 interaction effect} only varies with $\lambda$. The loglikelihood function component does not vary with $\lambda$, implying that $\ell_\lambda(\hat{\bm\beta}_\lambda;\hat\phi_\lambda)$ strictly decreases in $\lambda$. We can find cutting points $\lambda_0,\dots,\lambda_{v-1}$ to almost surely satisfy the requirement. Because strong hierarchy is required, the change of $|\hat\alpha_\lambda|$ can be greater than $1$ around the cutting points, implying that $v\le p\wedge n$. If $\lambda$ varies within $[\lambda_{s-1},\lambda_s)$, then $\ell(\hat{\bm\beta}_\lambda;\hat\phi_\lambda)$ does not vary implying that $\hat\alpha_\lambda$ does not vary. We conclude. \qed

\noindent
{\bf Proof of Theorem~\ref{thm:solution of lambda}.} The conclusion is directly implied by Lemma~\ref{lem:monotonicity of GIC}.  \qed

\noindent
{\bf Proof of Corollary~\ref{cor:GIC and L0 penalty}.} The conclusion is obvious by Theorem~\ref{thm:solution of lambda}. \qed

\noindent
{\bf Proof of Lemma~\ref{lem:basic FCLT without the linear transformation}.} The conclusion is directly implied by Corollary 1 of~\cite{herrndorf1984}. \qed

\noindent
{\bf Proof of Theorem~\ref{thm:asymptotic distributions of the terms}.} We combine the FCLT given by Lemma~\ref{lem:basic FCLT without the linear transformation} with approximation
$$\sqrt{n\over\phi^*}(\hat{\bm\beta}_\alpha-\hat{\bm\beta}_{\alpha^*}^\alpha)= {\cal I}_\alpha^{-1}({\bf I}-{\cal I}_\alpha{\cal Q}_{\alpha}) {1\over\sqrt{n}}\dot\ell({\bm\beta}_{\alpha^*}^{\alpha}) +o_p(1)$$
given by~\cite[P. 146]{ferguson1996}. Because $\dot\ell({\bm\beta}_{\alpha^*}^\alpha)$ is a linear transformation of ${\bm y}-{\bm\mu}^*$, we obtain~\eqref{eq:asymptotic distribution of the MLE in the selected and true model} and~\eqref{eq:joint distribution of MLE of two models}. By the Continuous Mapping Theorem, we obtain~\eqref{eq:asymptotic distribution of the likelihood ratio given alpha}. \qed

\noindent
{\bf Proof of Lemma~\ref{lem:asymptotic distribution of the likelihood ratio statistics many models}.} By Lemma~\ref{lem:basic FCLT without the linear transformation}, we have
$$
\sqrt{\phi^*\over{n}}\left[\left(\begin{array}{c} \dot\ell({\bm\beta}_{\alpha_1}) \cr \dot\ell({\bm\beta}_{\alpha_2})\cr\end{array} \right)-\left(\begin{array}{c} {\bf X}_{\alpha_1}^\top({\bm\mu}^*-{\bm\mu}_\alpha)  \cr {\bf X}_{\alpha_2}^\top({\bm\mu}^*-{\bm\mu}_\alpha) \cr\end{array}  \right)\right]\rightsquigarrow {\cal N}\left[\left(\begin{array}{c}{\bf 0}\cr{\bf 0} \end{array}   \right),  \left(\begin{array}{cc} {\cal I}_{\alpha_1} & {\cal I}_{\alpha_1\alpha_2} \cr {\cal I}_{\alpha_2\alpha_1} & {\cal I}_{\alpha_2}\cr   \end{array}\right)  \right]
$$
for any distinct $\alpha_1,\alpha_2\in\Omega_+\cup\Omega_*$. Thus, ${\bf A}=({\cal I}_{\alpha_{i}\alpha_{j}})_{i,j\in\{1,\dots,m\}}$
is a valid variance-covariance matrix. Let ${\bf B}={\rm diag}({\cal I}_{\alpha_1}^{-1}({\bf I}-{\cal I}_{\alpha_1}{\cal Q}_{\alpha_1}),\dots, {\cal I}_{\alpha_k}^{-1}({\bf I}-{\cal I}_{\alpha_k}{\cal Q}_{\alpha_m}))$. Because ${\bf B}{\bf A}{\bf B}=(  {\bm\Sigma}_{\alpha_i\alpha_j} )_{i,j\in\{1,\dots,m\}}$ is a valid covariance-covariance matrix, we conclude that ${\bm z}_{\alpha_1},\dots,{\bm z}_{\alpha_m}$ exist and are unique by the Kolmogorov existence theorem~\cite[P. 483]{billingsley1995}. \qed

\noindent
{\bf Proof of Corollary~\ref{cor:joint limit distribution of likelihood ratio}.} The conclusion can be directly implied by the implementation of Continuous Mapping Theorem to Lemma~\ref{lem:asymptotic distribution of the likelihood ratio statistics many models}. \qed

\noindent
{\bf Proof of Lemma~\ref{lem:the linear term}.}  Let $a_n=(c_1n)^{1/2}\|{\bm\beta}_{\alpha}-{\bm\beta}^*\|$ and $b_n=(c_2n)^{1/2}\|{\bm\beta}_{\alpha}-{\bm\beta}^*\|$.  By $T_2({\bm\beta}_{\alpha},\check{\bm\beta}_{\tilde\alpha})\ge c_1n\|{\bm\beta}_{\alpha}-{\bm\beta}^*\|^2=c_1n^{2\epsilon}(n^{1/2-\epsilon}\| {\bm\beta}_{\alpha}-{\bm\beta}^*\|)^2\rightarrow\infty$ as $n\rightarrow\infty$, we obtain $a_n^2\le T_2({\bm\beta}_{\alpha},\check{\bm\beta}_{\tilde\alpha})\le b_n^2$ for any ${\bm\beta}_{\alpha}$. By ${\rm E}_{{\bm\beta}^*}[a_n^{-1}T_1({\bm\beta}_{\alpha})]=a_n^{-1}{\rm E}_{{\bm\beta}^*}[T_1({\bm\beta}_{\alpha})]=0$ and ${\rm V}_{{\bm\beta}^*}[a_n^{-1}T_1({\bm\beta}_{\alpha})]=a_n^{-2}{\rm V}_{{\bm\beta}^*}[T_1({\bm\beta}_{\alpha})]=a_n^{-2}T_2({\bm\beta}_{\alpha})\in[1,(c_2/c_1)^{1/2}]$, we obtain $a_nT_1({\bm\beta}_{\alpha})=O_p(1)$. By $a_n\rightarrow\infty$ as $n\rightarrow\infty$, we draw the conclusion. \qed

\noindent
{\bf Proof of Theorem~\ref{thm:consistency under the low-dimensional setting}.} Because $|\Omega|$ is bounded, we can directly implement Corollary~\ref{cor:joint limit distribution of likelihood ratio} to the entire $\Omega_+$. By $\Omega_+=\bigcup_{r=1}^{p-|\alpha^*|} B_r$ with $B_r=\{\alpha\supset\alpha^*:|\alpha|-|\alpha^*|=r\}$ for $r=1,\dots,p(p+1)/2-|\alpha^*|$, we have
$$\eqalign{
&\lim_{n\rightarrow\infty}{\rm Pr}\{\hat\alpha_\lambda\in\Omega_+\}\cr
=&\lim_{n\rightarrow\infty}{\rm Pr}\{\max_{\alpha\in\Omega_+} 2[\ell(\hat{\bm\beta}_{\alpha};\hat\phi_{\alpha})-\ell(\hat{\bm\beta}_{\alpha^*}^{\alpha};\hat\phi_{\alpha^*}^{\alpha})]\ge{n}\lambda(|\alpha|-|\alpha^*|)\}\cr
\le & \lim_{n\rightarrow\infty} \sum_{r=1}^{p(p+1)/2-|\alpha^*|}  {\rm Pr}\{\max_{\alpha\in B_r}  q_\alpha\ge rn\lambda\}\cr
\le &\lim_{n\rightarrow\infty} {p(p+1)\over 2}{\rm Pr}\{\chi_1^2\ge n\lambda\}+\lim_{n\rightarrow\infty}  \sum_{r=2}^{p(p+1)/2-|\alpha^*|} \left[{p(p+1)\over2}-|\alpha^*|\right]^r {\rm Pr}\{\chi_r^2\ge rn\lambda\}.
}$$
By the upper tail probability formula of the ${\cal N}(0,1)$, we have ${\rm Pr}\{\chi_1^2\ge n\lambda\}\le(n\lambda)^{-1/2} e^{-\kappa/2}$ if $n\lambda$ is sufficiently large. If we choose $\lambda=n^{-1}\log{n}$, then the first term is $0$. We only need to evaluate the second term. By~\eqref{eq:upper tail chi-square distribution}
 the upper tail probability of the $\chi^2$-distribution given by~\cite{inglot2010} as
\begin{equation}
\label{eq:upper tail chi-square distribution}
{\rm Pr}\{\chi_r^2\ge u\}\le {u\over\sqrt{\pi}(u-r+2)}\exp\left\{-{1\over 2}[u-r-(r-2)\log(u/r)+\log{r}]\right\}
\end{equation}
for any $r\ge 2$ and $u\ge r-2$, if we choose $\lambda=n^{-1}\log{n}$, then
$$\eqalign{
&\lim_{n\rightarrow\infty}  \sum_{r=2}^{p(p+1)/2-|\alpha^*|} \left[{p(p+1)\over 2}-|\alpha^*|\right]^r {\rm Pr}\{\chi_r^2\ge {n\lambda} r\}\cr
\le & \lim_{n\rightarrow\infty} \sum_{r=2}^{p(p+1)/2-|\alpha^*|}{p^r(p+1)^r\over2^r\sqrt{\pi}}{n\lambda\over{n}\lambda-1}\exp\left\{-{1\over 2}[(n\lambda-1)r-(r-1)\log(n\lambda)+\log{r}] \right\}\cr
=&0,\cr
}$$
implying that $\hat\alpha_\lambda$ is not overfitted in probability. For any $\alpha\in\Omega_-$, by the Chebyshev inequality, we obtain $n^{-(1-v)}\ell(\hat{\bm\beta}_\alpha;\hat\phi_\alpha)-{\rm E }[n^{-(1-v)}\ell(\hat{\bm\beta}_\alpha;\hat\phi_\alpha)]\stackrel{P}\rightarrow 0$ for any $v<1/2$. By the connection between ${\rm E}[\ell(\hat{\bm\beta}_{\alpha^*};\hat\phi_{\alpha^*})]$ and ${\rm KL}(\hat{\bm\beta}_\alpha)$, we obtain $(n^{-(1-v)}/\phi)[\ell(\hat{\bm\beta}_\alpha;\phi)-\ell(\hat{\bm\beta}_{\alpha^*};\phi)]+n^{-(1-v)}{\rm KL}(\hat{\bm\beta}_\alpha)\stackrel{P}\rightarrow 0$ for any $\phi>0$, leading to $\lim_{n\rightarrow\infty}{\rm Pr}\{n^{-(1-v)}[\ell(\hat{\bm\beta}_\alpha;\phi)-\ell(\hat{\bm\beta}_{\alpha^*};\phi)]\le -n^{v}\delta_n\}=1$. Combining this with (ii), we obtain $[\ell(\hat{\bm\beta}_\alpha;\phi)-\ell(\hat{\bm\beta}_{\alpha^*};\phi)]/\log{n}\stackrel{P}\rightarrow-\infty$. Because $\hat{\bm\beta}_\alpha$ and $\hat{\bm\beta}_{\alpha^*}$ do not depend on $\phi$, by consistency of the MLEs of $\phi$ in both cases, we obtain ${\rm Pr}\{\hat\alpha\in\Omega_-\}=0$. Thus, $\hat\alpha_\lambda\stackrel{P}\rightarrow\alpha^*$. Because $\hat{\bm\beta}_\lambda$ and $\hat{\phi}_\lambda$ are identical to the MLEs of ${\bm\beta}$ and $\phi$ under $\hat\alpha_{\lambda}$, we draw the oracle properties.  \qed

\noindent
{\bf Proof of Lemma~\ref{lem:FCLT under high-dimensional setting}.} Let $\tilde F_{nj,\alpha}$ be the true CDF, $\tilde \Phi_{j,\alpha}$ be the limiting distribtion of the $j$th component of $n^{-1/2}\dot\ell({\bm\beta}_{\alpha^*}^\alpha;\phi^*)$, and $F_{nj,\alpha}$ be the true CDF and $\Phi_{j,\alpha}$ be the limiting distribtion of the $j$th component of $\sqrt{n}(\hat{\bm\beta}_\alpha-\hat{\bm\beta}_{\alpha^*}^\alpha)$. Then, $\sqrt{n/\phi^*}\|\tilde F_{nj,\alpha}-\tilde \Phi_{j,\alpha}\|_\infty$ is uniformly bounded in probability by the Berry-Esseen Theorem~\cite{shevtsova2013}. Because the maximum eigenvalue of ${\cal I}_\alpha^{-1}({\bf I}-{\cal L}_\alpha{\cal Q}_\alpha)$ is uniformly bounded, $\sqrt{n/K}\|F_{nj,\alpha}-\Phi_{j,\alpha}\|_\infty$ is uniformly bounded, implying that $\sup_{\alpha\in\Omega^+}\| F_{nj,\alpha}- \Phi_{j,\alpha}\|_\infty\stackrel{P}\rightarrow 0$. The maximum inequality is satisfied.\qed

\noindent
{\bf Proof of Theorem~\ref{thm:consistency of ALRSIS and ASSIS}.} By~\eqref{eq:KL for underfitted}, for any $j'\in\bar\alpha_M\cap\bar\alpha_M^*$ and $j\in\bar\alpha_M\cap\alpha_M^*$,  there is $\lim_{n\rightarrow\infty}[{\rm E}(aG_{j|\alpha}-aG_{j'|\alpha})-n\delta_n]\ge 0$.  By $G_\alpha^2- G_{\alpha\cup\{(j',k')\}}^2-{\rm E}(G_\alpha^2- G_{\alpha\cup\{(j',k')\}}^2)\rightsquigarrow\chi_1^2$, for any  $j\in\bar\alpha_M\cap\alpha_M^*$, we have
$$\eqalign{
&\lim_{n\rightarrow\infty}{\rm Pr}\{  \max_{j'\in\bar\alpha_M\cap\bar\alpha_M^*} aG_{j'|\alpha}  \ge aG_{j|\alpha} \}\cr
\le &\lim_{n\rightarrow\infty} \sum_{j'\in\bar\alpha_M\cap\bar\alpha_M^*}{\rm Pr}\{aG_{j'|\alpha} \ge aG_{j|\alpha} \}\cr
\le & \lim_{n\rightarrow\infty} \sum_{j'\in\bar\alpha_M\cap\bar\alpha_M^*}\sum_{k'\in\bar\alpha_M\cup\{0\}} {\rm Pr}\{ G_\alpha^2- G_{\alpha\cup\{(j',k')\}}^2  \ge  aG_{j|\alpha}\}\cr
\le & \lim_{n\rightarrow\infty}p^2 {\rm Pr}(\chi_1^2\ge n\delta_n).\cr
}$$
Note that $p^2< e^{n^2\delta_n^2/2}$ under (a) and (b), respectively. By  $\lim_{x\rightarrow\infty} e^{x^2/2}{\rm Pr}(\chi_1^2\ge x)=0$ and the relationship between the likelihood ratio and the score statistics, we conclude. \qed 

\noindent
{\bf Proof of Theorem~\ref{thm:goodness of the fit statistic of the true model}}. Because $p>n$, we can construct saturated $\check\alpha$ with $n-1$ main effects and interactions  such that it satisfies ${\bm y}=g^{-1}({\bf X}\hat{\bm\beta}_{\check\alpha})$ and $\ell(\hat{\bm\beta}_{\check\alpha};\hat\phi_{\check\alpha})\ge\ell(\hat{\bm\beta}_{\alpha};\hat\phi_{\alpha})$ for any $\alpha\in\Omega$ implying that it has the largest possible likelihood value from~\eqref{eq:high-dimensional GLM with interaction effects}, where $\hat\phi_{\check\alpha}$ can be arbitrary. By $\ell(\hat{\bm\beta}_{\hat\alpha_\lambda};\hat\phi_{\hat\alpha_\lambda})-(n\lambda/2)|\hat\alpha_\lambda|>\ell(\hat{\bm\beta}_{\alpha^*};\hat\phi_{\alpha^*})-(n\lambda/2)|\alpha^*|>\ell(\hat{\bm\beta}_{\alpha^*};\phi^*)-(n\lambda/2)|\alpha^*|$, we obtain $2[\ell(\hat{\bm\beta}_{\check\alpha};\phi^*)-\ell(\hat{\bm\beta}_{\alpha^*};\phi^*)]>2[\ell(\hat{\bm\beta}_{\check\alpha};\hat\phi_{\check\alpha})-\ell(\hat{\bm\beta}_{\hat\alpha};\hat\phi_{\hat\alpha_\lambda})]+n\lambda(|\hat\alpha_\lambda|-|\alpha^*|)\ge n\lambda(|\hat\alpha_\lambda|-|\alpha^*|)$. By $2[\ell(\hat{\bm\beta}_{\check\alpha};\phi^*)-\ell(\hat{\bm\beta}_{\alpha^*};\phi^*)]\rightsquigarrow\chi_{a}^2$ where $a=n-|\alpha^*|-1$, we have $\lim_{n\rightarrow\infty}P\{2[\ell(\hat{\bm\beta}_{\check\alpha};\phi^*)-\ell(\hat{\bm\beta}_{\alpha^*};\phi^*)]\ge a(1+\epsilon)\log(a)\}=\lim_{n\rightarrow\infty} P\{\chi_{a}^2\ge a(1+\epsilon)\log{a}\}$ for any $\epsilon>0$. By~\eqref{eq:upper tail chi-square distribution}, we conclude that it is less than $\lim_{\nu\rightarrow\infty} \pi^{-1/2}\exp\{-(1/2)[ a(1+\epsilon)\log{a}-a-(a-2) \log(1+\epsilon)-(a-2)\log\!\log{a} +\log{a} ]\}$, which is less than $\lim_{\nu\rightarrow\infty} \pi^{-1/2}\exp(-a\epsilon\log{a}/4)=0$ by $\lim_{n\rightarrow\infty} (a/n)=1$. By $a<n$ and $n\lambda>(4+\epsilon)\log{p}$, we obtian $\lim_{n\rightarrow\infty} {\rm Pr}[|\hat\alpha_\lambda|<|\alpha^*|+a(1+\epsilon)\log(a)/(n\lambda)]=1$, leading to $\lim_{n\rightarrow\infty} {\rm Pr}\{|\hat\alpha_\lambda|<|\alpha^*|+(1+\epsilon){n}\log(n)/[(4+\epsilon)\log{p}]\}=1$. By $(1+\epsilon)/(4+\epsilon)<1/4$ for any $\epsilon>0$, we obtain $\lim_{n\rightarrow\infty} {\rm Pr}[|\hat\alpha_\lambda|<|\alpha^*|+{n}\log(n)/(4\log{p})]=1$. We obtain the first conclusion. The remaining two conclusions can be easily derived by adjusting $(1+\epsilon)/(4+\epsilon)$ to a value less than $1/4$ in the evaluation of the limit. \qed

\noindent
{\bf Proof of Theorem~\ref{thm:consistency when p polynomially increases with n}.} Let $a=p(p+1)/2$. By $\Omega_{+}=\bigcup_{r=1}^{K-s^*} B_r$ with $B_r=\{\alpha\supset\alpha^*:|\alpha|-s^*=r\}$ for $r=1,\dots,p-s^*$ satisfying $|B_r|={a-|\alpha^*|\choose r}\le {a\choose r}\le a^r$ and Lemma~\ref{lem:FCLT under high-dimensional setting}, we have ${\rm Pr}(\hat\alpha_\lambda\in\Omega_+)={\rm Pr}\{\max_{\alpha\in \Omega_{+}} 2[\ell(\hat{\bm\beta}_\alpha;\hat\phi_\alpha)-\ell(\hat{\bm\beta}_{\alpha^*}^\alpha;\hat\phi_{\alpha^*})\}\ge n\lambda(|\alpha|-|\alpha^*|)]={\rm Pr}[\max_{\alpha\in \Omega_{+}} q_\alpha\ge{n}\lambda(|\alpha|-|\alpha^*|)]+o_p(1)\le \sum_{r=1}^{K-|\alpha^*|}a^r  {\rm Pr}(\chi_{r}^2 \ge n\lambda r)+o_p(1)$. Because $a{\rm Pr}(\chi_1^2\ge 4\log{p})=o_p(1)$, we can ignore the first term in the summation, leading to that ${\rm Pr}(\hat\alpha_\lambda\in\Omega_+)\le \sum_{r=2}^{K-|\alpha^*|}a^r  {\rm Pr}(\chi_{r}^2 \ge n\lambda r)+o_p(1)$. After a few step of algebra based on~\eqref{eq:upper tail chi-square distribution}, we obtain ${\rm Pr}(\hat\alpha_\lambda\in\Omega_+)\le \sum_{r=2}^{K-|\alpha^*|}\pi^{-1/2} \exp\{-(1/2)[n\lambda{r}-r(1+2\log{a})-(r-2)\log(n\lambda)+\log r]\}+o_p(1)$. If we choose $\lambda=[(4+\epsilon)/n]\log{p}$ for any $\epsilon>0$ or $\lambda=(1/n)\log{p}\log\!\log{n}$, then for sufficiently large $n$ we have $n\lambda{r}-r(1+2\log a)-(r-2)\log\kappa+\log r= r(\epsilon\log a-1)-(r-2)\log(2+\epsilon/2)-(r-2)\log\!\log a+\log r \ge  (r/2)\log a$, leading to $\lim_{n\rightarrow\infty}{\rm Pr}\{\hat\alpha\in\Omega_+\}=\lim_{n\rightarrow\infty}\sum_{r=2}^{K-|\alpha^*|}\pi^{-1/2}e^{-(r/4)\log{a}}=0$ under either (a) or (b). For any $\alpha\in\Omega_-$, we extend it to overfitted $\tilde\alpha=\alpha\cup\alpha^*$. We define $\chi^2$-statistic $q_{\tilde\alpha}$ by Corollary~\ref{cor:joint limit distribution of likelihood ratio}. By Lemma~\ref{lem:the linear term}, we have $\lim_{n\rightarrow\infty}{\rm Pr}\{2n^{-1}[\ell(\hat{\bm\beta}_{\alpha};\phi^*)-\ell({\bm\beta}^*;\phi^*)]\le q_{\tilde\alpha}-(1-\epsilon)\delta_n\}=1$ for any $\epsilon>0$. Let $\bar\Omega_-=\{\tilde\alpha: \tilde\alpha=\alpha\cup\alpha^*,\alpha\in\Omega_-\}$. Then, $\bar\Omega_-=\bigcup_{r=0}^{K-s^*} \bar B_r$, where $\bar B_r=\{\alpha\in\bar\Omega_-: |s^*\backslash\alpha|=r\}$. By Lemma~\ref{lem:FCLT under high-dimensional setting}, we have  $\lim_{n\rightarrow\infty} {\rm Pr}(\alpha\in\Omega_-)=\lim_{n\rightarrow\infty}{\rm Pr}\{\max_{\alpha\in\Omega_-}2[\ell(\hat{\bm\beta}_{\alpha};\hat\phi_\alpha)-\ell(\hat{\bm\beta}_{\alpha^*};\hat\phi_{\alpha^*})]\ge n\lambda(|\alpha|-|\alpha^*|)  \} \le \lim_{n\rightarrow\infty}  \sum_{r=2}^{K-|\alpha^*|} {\rm Pr}(\chi_r^2\ge {n}\lambda{r})$. By~\eqref{eq:upper tail chi-square distribution}, we obtain $\lim_{n\rightarrow\infty} {\rm Pr}(\alpha\in\Omega_-)\le  \lim_{n\rightarrow\infty}  \sum_{r=2}^{K-|\alpha^*|} \pi^{-1/2}a^r\exp\{-(1/2)[n\lambda{r}-r-(r-2)\log(n\lambda)+\log{r}]\}$. If $\lambda=[(4+\epsilon)/n]\log{p}$ for some $\epsilon>0$ or $\lambda=(1/n)\log{p}\log\!\log{n}$, then we have 
$\lim_{n\rightarrow\infty} {\rm Pr}\{\alpha\in\Omega_-\}\le \lim_{n\rightarrow\infty} \sum_{r=2}^{K-|\alpha^*|}\pi^{-1/2} \exp\{-(1/2)[\epsilon{r}\log{a}-r-(r-2)\log[(2+\epsilon/2)\log{a}]+\log{r}]\}$. If (a) or (b) of Theorem~\ref{thm:goodness of the fit statistic of the true model}, then $\log{p}\rightarrow\infty$ as $n\rightarrow\infty$. If $n$ is sufficiently large, then $n+(n-2)\log[(2+\epsilon/2)\log{a}]+\log{n}<(\epsilon/4) r\log{a}$, leading to $\lim_{n\rightarrow\infty} {\rm Pr}(\alpha\in\Omega_-) \le  \lim_{n\rightarrow\infty} \sum_{r=2}^{K-|\alpha^*|}\pi^{-1/2} \exp\{-(1/4)\epsilon{r}\log{a}\}=\lim_{n\rightarrow\infty} \sum_{r=2}^{K-|\alpha^*|}\pi^{-1/2} a^{-\epsilon{r}/4}=0$.
Because $\epsilon>0$ is arbitrary, we draw $\hat\alpha_{\lambda}\stackrel{P}\rightarrow \alpha^*$. The oracle properties can be straightforwardly obtained. \qed

\noindent
{\bf Proof of Corollary~\ref{cor:extension to the case when phi is not present}.} We draw the conclusions by simply treating $\hat\phi_\alpha=1$ as the estimator of $\phi$ for all $\alpha\in\Omega$. \qed

\noindent
{\bf Proof of Theorem~\ref{thm:extension to general high-dimensional statistical models}.} Because the conclusions can be proven by the methods used in the proofs of Theorems~\ref{thm:consistency under the low-dimensional setting} and~\ref{thm:consistency when p polynomially increases with n}, we omit the details. \qed

\noindent
{\bf Proof of Theorem~\ref{thm:local optimizers}.} We use the same method in the proof of Theorem~\ref{thm:consistency when p polynomially increases with n} to show the conclusion, because we only need the upper tail probability of the $\chi^2$-distribution and adjustment of the multiple testing problems for a family of likelihood ratio statistics. In the addition step, if $\alpha\not\subseteq\alpha^*$, then there exists either main effect ${\bm x}_j$ or interaction effect ${\bm x}_j\circ{\bm x}_k$ such that $j\in\alpha_M^*$ but $j\not\in\alpha_M$ or $(j,k)\in\alpha_I^*$ but $(j,k)\not\in\alpha_I$. By the method used in the proof of Theorem~\ref{thm:consistency when p polynomially increases with n}, we can show $P(\{\tilde\alpha=\alpha\cup\{j\}: j\not\in\alpha_M^*\})\rightarrow 0$ and $P(\{\tilde\alpha=\alpha\cup\{(j,k)\}: (j,k)\not\in\alpha_M^*\})\rightarrow 0$. If $\alpha\not\subseteq\alpha^*$, then there exists $j\in\alpha$ but $j\not\in\alpha^*$ or $(j,k)\in\alpha$ but $(j,k)\in\alpha^*$. Similarly, we can show $P(\{\tilde\alpha=\alpha\backslash\{j\}:j\in\alpha_M^*\})\rightarrow 0$ and $P(\{\tilde\alpha=\alpha\backslash\{(j,k)\}: j\in\alpha_M^*\})\rightarrow 0$. As strong hierarchy is satisfied by $\alpha^*$, we only need to consider $\tilde\alpha$ satisfying the SH restriction. The condition $\lim\!\inf_{n\rightarrow\infty}n^v\delta_n>0$ for some $v<1/2$ ensures that $n^{1/2}\beta_j^*$ for $j\in\alpha_M^*$ or $n^{1/2}\beta_{jk}^*$ for $(j,k)\in\alpha_I^*$ do not shrink to $0$ as $n\rightarrow\infty$, implying that remaining steps of the method can be applied. We apply those to the case when $\phi$ is present or absent, leading to the conclusion for HDGLMs. Repeat the remaining part. We conclude for general ultrahigh-dimensional statistical models. \qed

\noindent
{\bf Proof of Corollary~\ref{cor:finite step iterations}.} In each iteration, at least one main effect or interaction contained by $\alpha^*$ is added in probability. After $|\alpha^*|$ iterations, $\alpha^*$ is covered in probability.  After the main effects and interactions contained by $\alpha^*$ are included, none of them would be removed in probability.  Then, in each iteration, at least one main or interaction effect not contained by $\alpha^*$ is removed in probability. This can induce at most $|\alpha^*|^2$ more interactions to remove all main effects and interactions not contained by $\alpha^*$ in probability. We conclude. \qed


\begin{thebibliography}{}
\bibitem{barutfan2016}
Barut, E., Fan, J., and Verhasselt, A. (2016). Conditional sure independence screening. {\it Journal of the American Statistical Association}, {\bf 111}, 1266-1277.
\bibitem{bien2013}
Bien, J., Taylor, J., Tibshirani, R. (2013). A LASSO for hierarchical interactions. {\it Annals of Statistics}, {\bf 41}, 1111-1141.
\bibitem{billingsley1995}
Billingsley, P. (1995). {\it Probability and Measure}, Wiley, New York.
\bibitem{bertsimas2016}
Bertsmas, D., King, A., and Mazumder, R. (2016). Best subset selection via a modern optimization lens. {\it Annals of Statistics}, {\bf 44}, 813-852.
\bibitem{chipman1996}
Chipman, H. (1996). Bayesian variable selection with related predictors. {\it Canadian Journal of Statistics}, {\bf 24}, 17–36.
\bibitem{choi2010}
Choi, N.H., Li, W., and Zhu, J. (2010). Variable selection With the strong heredity constraint and its oracle property. {\it Journal of the American Statistical Association},{\bf 105}, 354-364,
\bibitem{cook1971}
Cook, S.A. (1971). The complexity of theorem-proving procedures. {\it Proceedings of the third annual ACM symposium on Theory of computing- STOC}, 151–158.
\bibitem{crama2005}
Crama, Y., Kolen, A. W., and Pesch, E. J. (2005). Local search in combinatorial optimization. {\it Artificial Neural Networks: An Introduction to ANN Theory and Practice}, 157-174.
\bibitem{dicker2013}
Dicker, L., Huang, B., and Lin, X. (2013). Variable selection and estimation with the seamless-$L_0$ penalty. {\it Statistica Sinica}, {\bf 23}, 929-962.
\bibitem{du1998}
Du, D. and Pardalos, P.M. (1998). {\it Handbook of Combinatorial Optimization}, Kluwer Adademic Publishers, Boston, MA.
\bibitem{fanlv2008}
Fan, J., and Lv, J. (2008). Sure independence screening for ultrahigh dimensional feature spce. {\it Journal of the Royal Statistical Society Series B}, {\bf 70}, 849-911.
\bibitem{fanli2001}
Fan, J. and Li, R. (2001). Variable selection via nonconcave penalized likelihood and its oracle properties. {\it Journal of the American Statistical Association}, {\bf 96}, 1348-1360.
\bibitem{fantang2013}
Fan, Y. and Tang, C.Y. (2013). Tuning parameter selection in high dimensional penalized likelihood. {\it Journal of Royal Statistical Society Series B}, {\bf 75}, 531-552.
\bibitem{ferguson1996}
Ferguson, T.S. (1996). {\it A Course in Large Sample Theory}, CRC Press, Boca Raton, Florida.
\bibitem{hallxue2014}
Hall, P., and Xue, J.H. (2014). On selecting interacting features from high-dimensional data. {\it Computational Statistic and Data Analysis}, {\bf 71}, 694-708.
\bibitem{hamada1992}
Hamada, M., and Wu, C. F. J. (1992). Analysis of designed experiments with complex aliasing. {\it Journal of Quality Technology}, {\bf 24}, 130-137.
\bibitem{hao2018}
Hao, N., Feng, Y., and Zhang, H.H. (2018). Model selection for high-dimensional quadratic regression via regularization. {\it  Journal of the American Statistical Association}, {\bf 113}, 615-625.
\bibitem{hazimeh2020}
Hazimeh, H., and Mazumder, R. (2020). Learning hierarchical interactions at scale: A convex optimization approach. In {\it International Conference on Artificial Intelligence and Statistics} (pp. 1833-1843). PMLR.
\bibitem{herrndorf1984}
Herrnodorf, H. (1984). A functional central limit theorem for weakly dependent sequences of random variables. {\it Annals of Probability}, {\bf 12}, 141-153.
\bibitem{huang2018}
Huang, J., Jiao, Y., Liu, Y., and Lu, X. (2018). A constructive approach to $\ell_0$ penalized regression. {\it Journal of Machine Learning Research}, {\bf 19}, 403-439.
\bibitem{inglot2010}
Inglot, T. (2010). Inequalities for quantiles of the chi-square distribution. {\it Probability and Mathematical Statistics}, {\bf 30}, 339-351. 
\bibitem{kim2021}
Kim, J., Zhu, H., Wang, X., and Do, K.A. (2021). Scalable network estimation with $L_0$ penalty. {\it Statistical Analysis and Data Mining}, {\bf 14}, 18-30.
\bibitem{kongli2017}
Kong, Y., Li, D., Fan, Y., and Lv, J. (2017). Interaction pursuit in high-dimensional multi-response regression via distance correlation. {\it Annals of Statistics}, {\bf 45}, 897-922.
\bibitem{levin1973}
Levin, L. (1973). A survey of Russian approaches to perebor (brute-force searches) algorithms. {\it Annals of the History of Computing},  {\bf 6}, 384–400.
\bibitem{li2018}
Li, X., Xie, S., Zeng, D., Wang, Y. (2018). Efficient $\ell_0$-norm feature selection based on augmented and penalized minimization. {\it Statistical Methods}, {\bf 38}, 473-486.
\bibitem{li2019}
Li, Y. and Liu, J.S. (2019). Robust variable and interaction selection for logistic regression and general index models. {\it Journal of the American Statistical Association}, {\bf 114}, 271-286.
\bibitem{liu2016}
Liu, Z., and Li, G. (2016). Efficient regularized regression with $L_0$ penalty for variable selection and network construction. {\it Computational and Mathematical Methods in Medicine}, 3456153.
\bibitem{mccullagh2002}
McCullagh, P. (2002). What is a statistical model? {\it Annals of Statistics}, {\bf 30}, 1225–1267
\bibitem{orlin2004}
Orlin, J. B., Punnen, A. P., and Schulz, A. S. (2004). Approximate local search in combinatorial optimization. {\it SIAM Journal on Computing}, {\bf 33}, 1201-1214.
\bibitem{papadimitriou1998}
Papadimitriou, C. H., and Steiglitz, K. (1998). {\it Combinatorial optimization: algorithms and complexity}. Dover Publications, Mineola, New York, USA.
\bibitem{radchenko2010}
Radchenko, P., and James, G.M. (2010). Variable selection using adaptive nonlinear interaction structures in high dimensions. {\it Journal of the American Statistical Association}, {\bf 105}, 1541-1553.
\bibitem{she2018}
She, Y., Wang, Z., and Jiang, H. (2018). Group regularized estimation under structural hierarchy. {\it Journal of the American Statistical Association}, {\bf 113}, 445-454.
\bibitem{shevtsova2013}
Shevtsova, I.G. (2013). On the absolute constants in the Berry–Esseen inequality and its structural and nonuniform improvements. {\it Informatika i Ee Primeneniya (Informatics and Its Applications}, {\bf 7}, 124-125. 
\bibitem{singh2002}
Singh, D., Febbo, P.G., Ross, K., Jackson, D., Manalo, J., Ladd, C., Tamayo, P., Renshaw, A.A., D'Amico, A.V., Richie, J.P., Lander, E.S., Loda, M., Kantoff, P.W., Golub, T.R, and Sellers, W.R. (2002). Gene expression correlates of clinical prostate cancer behavior. {\it Cancer Cell}, {\bf 1}, 203-209.
\bibitem{tibshirani1996}
Tibshirani, R.J. (1996). Regression shrinkage and selection via the LASSO. {\it Journal of the Royal Statistical Society Series B}, {\bf 58}, 267-288.
\bibitem{vandervaart1998}
van der Vaart, A.W. (1998). {\it Asymptotic Statistics}, Cambridge University Press, Cambridge, UK.
\bibitem{wang2024}
Wang, C., Chen, H., and Jiang, B. (2024). HiQR: an efficient algorithm for high-dimensional quadratic regression with penalties. {\it Computational Statistics and Data Analysis}, {\bf 192}, 107904.
\bibitem{wangkim2013}
Wang, L., Kim, Y., Li, R. (2013). Calibrating nonconvex penalized regression in ultra-high dimension. {\it Annals of Statistics}, {\bf 41}, 2505-2536.
\bibitem{yu2023}
Yu, G., Bien, J., and Tibshirani, R. (2023). Reluctant interaction modeling. Preprint, arXiv: 1907.08414v3.
\bibitem{zhang2010}
Zhang, C.H. (2010). Nearly unbiased variable selection under minimax concave penalty. {\it Annals of Statistics}, {\bf 38}, 894-942.
\bibitem{zhang2024}
Zhang, T. (2024). Aggregated sure independence screening for variable selection with interaction structures. ArXiv: 2407:03558.
\bibitem{zhangli2010}
Zhang, Y., Li, R., and Tsai,, C. (2010). Regularization parameter selections via generalized information criterion. {\it Journal of the American Statistical Association}, {\bf 105}, 312-323.
\end{thebibliography}
\end{document}